\documentclass[aps,twocolumn,showpacs,superscriptaddress,amsmath,amssymb,amsfonts,floatfix,longbibliography]{revtex4-1}

\usepackage{graphicx}
\usepackage{dcolumn}
\usepackage{bm}
\usepackage[dvipsnames]{xcolor}
\usepackage{subfigure}
\usepackage{wrapfig}
\usepackage{cancel}
\usepackage{color}
\usepackage{textcomp}
\usepackage{xfrac}
\usepackage{amsmath}
\usepackage{mathtools}
\usepackage{natbib,hyperref}
\usepackage{amsfonts}
\usepackage{amssymb}
\usepackage{array}

\hypersetup{
	colorlinks=true, 
	citecolor=blue,  
}

\newcommand{\rev}[1]{\textcolor{black}{#1}} 
 
\newcommand{\LRG}{LaRhGe$_3$}

\newcommand{\Cp}{$C_P$}
\newcommand{\Ce}{$C_{\rm{el}}$}
\newcommand{\Tc}{$T_{\text c}$}

\newcommand{\mHct}{$B_{\rm{c2}}(0)$}

\newcommand{\mHc}{$B_{\rm{c}}(0)$}
\newcommand{\muSR}{$\mu \rm{SR}$}
\newcommand{\Ifmm}{$I4mm$}

\begin{document}


\title{Discovery of Superconductivity and Electron-Phonon Drag in \\ the Non-Centrosymmetric Weyl Semimetal \LRG }

\author{Mohamed Oudah}
    \email[Email: ]{mohamed.oudah@ubc.ca}
    \affiliation{Stewart Blusson Quantum Matter Institute, University of British Columbia, Vancouver, BC V6T 1Z4, Canada}

\author{Hsiang-Hsi Kung}
    \affiliation{Stewart Blusson Quantum Matter Institute, University of British Columbia, Vancouver, BC V6T 1Z4, Canada}
    
\author{Samikshya Sahu}%
    \affiliation{Stewart Blusson Quantum Matter Institute, University of British Columbia, Vancouver, BC V6T 1Z4, Canada}
    \affiliation{Department of Physics \& Astronomy, University of British Columbia, Vancouver, BC V6T 1Z1, Canada}

\author{Niclas Heinsdorf}
    \affiliation{Stewart Blusson Quantum Matter Institute, University of British Columbia, Vancouver, BC V6T 1Z4, Canada}
    \affiliation{Department of Physics \& Astronomy, University of British Columbia, Vancouver, BC V6T 1Z1, Canada}
    \affiliation{Max Planck Institute for Solid State Research, Heisenbergstrasse 1, 70569 Stuttgart, Germany}

\author{Armin Schulz}
    \affiliation{Max Planck Institute for Solid State Research, Heisenbergstrasse 1, 70569 Stuttgart, Germany}
    
\author{Kai Philippi}
    \affiliation{Max Planck Institute for Solid State Research, Heisenbergstrasse 1, 70569 Stuttgart, Germany}
    
\author{Marta-Villa De Toro Sanchez}
    \affiliation{TRIUMF, Vancouver, British Columbia, V6T 2A3 Canada}

\author{Yipeng Cai}%
    \affiliation{TRIUMF, Vancouver, British Columbia, V6T 2A3 Canada}
    
\author{Kenji Kojima}
    \affiliation{Stewart Blusson Quantum Matter Institute, University of British Columbia, Vancouver, BC V6T 1Z4, Canada}
    \affiliation{TRIUMF, Vancouver, British Columbia, V6T 2A3 Canada}

\author{Andreas P. Schnyder}
    \affiliation{Max Planck Institute for Solid State Research, Heisenbergstrasse 1, 70569 Stuttgart, Germany}

\author{Hidenori Takagi}
    \affiliation{Max Planck Institute for Solid State Research, Heisenbergstrasse 1, 70569 Stuttgart, Germany}

\author{Bernhard Keimer}
    \affiliation{Max Planck Institute for Solid State Research, Heisenbergstrasse 1, 70569 Stuttgart, Germany}

\author{Doug A. Bonn}%
    \affiliation{Stewart Blusson Quantum Matter Institute, University of British Columbia, Vancouver, BC V6T 1Z4, Canada}
    \affiliation{Department of Physics \& Astronomy, University of British Columbia, Vancouver, BC V6T 1Z1, Canada}

\author{Alannah M. Hallas}
      \email[Email: ]{alannah.hallas@ubc.ca}
    \affiliation{Stewart Blusson Quantum Matter Institute, University of British Columbia, Vancouver, BC V6T 1Z4, Canada}
    \affiliation{Department of Physics \& Astronomy, University of British Columbia, Vancouver, BC V6T 1Z1, Canada}
    \affiliation{Canadian Institute for Advanced Research, Toronto, Ontario, Canada M5G 1M1}


\date{\today}

\begin{abstract}
\rev{We present an exploration of the effect of electron-phonon coupling and broken inversion symmetry on the electronic and thermal properties of the semimetal \LRG. Our transport measurements reveal evidence for electron-hole compensation at low temperatures, resulting in a large magnetoresistance of 3000\% at 1.8~K and 14~T. The carrier concentration is on the order of $10^{21}\rm{/cm}^3$ with high carrier mobilities of $2000~\rm{cm}^2/\rm{Vs}$. When coupled to our theoretical demonstration of symmetry-protected \textit{almost movable} Weyl nodal lines, we conclude that \LRG\ supports a Weyl semimetallic state. We discover superconductivity in this compound with a \Tc\ of 0.39(1)~K and \mHc\ of 2.2(1)~mT, with evidence from specific heat and transverse-field muon spin relaxation. 
We find an exponential dependence in the normal state electrical resistivity below $\sim50$~K, while Seebeck coefficient and thermal conductivity measurements each reveal a prominent peak at low temperatures, indicative of strong electron-phonon interactions. To this end, we examine the temperature-dependent Raman spectra of \LRG\ and find that the lifetime of the lowest energy $A_1$ phonon is dominated by phonon-electron scattering instead of anharmonic decay. We conclude that \LRG\ has strong electron-phonon coupling in the normal state, while the superconductivity emerges from weak electron-phonon coupling. These results open up the investigation of electron-phonon interactions in the normal state of superconducting non-centrosymmetric Weyl semimetals.}


\end{abstract}

\maketitle

\section{Introduction}

Semimetals provide a fertile ground for investigating the interplay between topology, superconductivity, and electron-phonon interactions. Starting from the discovery of a Dirac semimetallic state in graphene~\cite{novoselov2005two,geim2007rise}, the study of semimetals has grown to include ever more exotic flavors of topological states in materials with Dirac or Weyl crossings in their band structures~\cite{burkov2016topological,yan2017topological,armitage2018weyl}. Many topological semimetals are simultaneously found to exhibit remarkable transport properties including extremely large and non-saturating magnetoresistances and ultrahigh carrier mobilities~\cite{ali2014large,shekhar2015extremely,kumar2017extremely}. In certain semimetals, these transport properties have been definitively linked with strong electron-phonon interactions, including WP$_2$~\cite{yao2019observation,Burch2021PRX}, PtSn$_4$~\cite{wu2016dirac,fu2020largely}, and NbGe$_2$~\cite{yang2021evidence,emmanouilidou2020fermiology}, all of which also manifest interesting topological states.

One particularly interesting class of semimetals are those with broken inversion symmetry. 
In metals, broken inversion symmetry leads to electronic and magnetic couplings that are forbidden in centrosymmetric materials, thereby promoting exotic states. One such example is the effect of antisymmetric SOC on superconductivity. ASOC results in the splitting of spin-degenerate bands, where the energy splitting is proportional to the degree of SOC of the constituent elements. When the ASOC is sufficiently strong, a mixture of spin singlet and spin triplet pairings will occur yielding an unconventional superconducting state~\cite{bauer_exp_PRL_04,bauer_book_review,Smidman_Agterberg_review_2017,yip_review_2014,Schnyder_review_2015}.

The rich physics associated with broken inversion symmetry has led to significant interest in the $RTX_3$ ($R$ = rare earth, $T$ = transition metal, $X = \rm{Ge, Si, Sn}$) family of materials, which crystallize in the non-centrosymmetric \Ifmm\ space group. Materials in this family have been known to exhibit heavy fermion behavior, complex magnetic states, and unconventional superconductivity~\cite{kimura2005pressure,sugitani2006pressure,okuda2007magnetic,kimura2007normal,settai2007recent,measson2009magnetic,kawai2008split,eguchi2011crystallographic,isobe2014srausi3}. They have also attracted attention in light of their ASOC split Fermi-surfaces~\cite{kawai2008split}. It has been demonstrated in this class of materials that the transition element $T$ plays a crucial role in the splitting of the Fermi-surface, while the $R$ and $X$ elements play a much smaller role~\cite{kawai2008split}.

\begin{figure*}[t!]
\centering
\includegraphics[width=16.5cm]{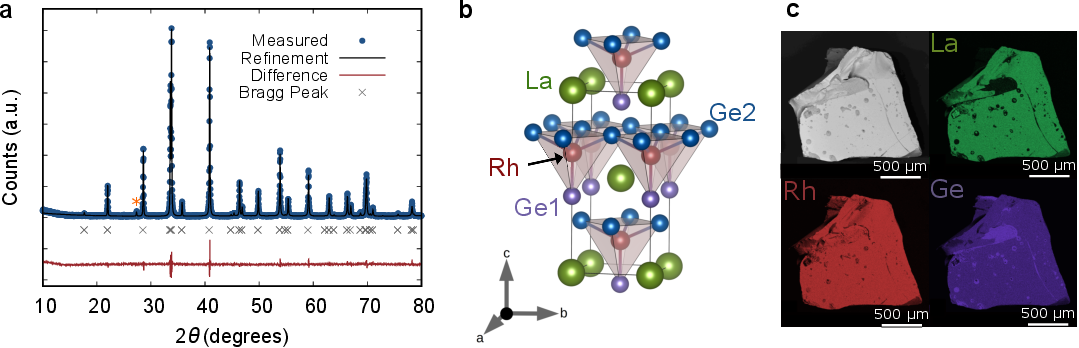}
\caption{Structural characterization of \LRG. (a) Rietveld refinement of the powder X-ray diffraction pattern of \LRG\ measured at room temperature. The Bragg peak positions for the $I4mm$ space group are indicated by the grey crosses. \rev{A small germanium impurity} (1.3\% by weight) peak is marked with an orange asterisk. (b) The non-centrosymmetric body-centered tetragonal crystal structure of \LRG, where La is shown in green, Rh in red, and the two Ge sites are shown in purple and blue, respectively. The broken inversion symmetry is most easily appreciated by considering the positions of Rh within the unit cell. (c) Scanning electron microscopy image of a single crystal of \LRG\ (top left) and elemental mapping on the same crystal for La (top right), Rh (bottom left), and Ge (bottom right). Intensity of colors in the elemental mapping corresponds to the intensity of peaks related to each element, showing uniform distribution of the elements across the crystal. 
}
\label{Fig1}
\end{figure*}

Here, we report a comprehensive characterization of one member of the non-centrosymmetric $RTX_3$ family, LaRhGe$_3$, which has significant antisymmetric SOC due to the presence of Rh on the $T$ site. \rev{We synthesize single crystals} of LaRhGe$_3$ using a self-flux method for the first time and characterize its low temperature electronic properties via transport, specific heat, muon spin relaxation, and Raman spectroscopy. This ensemble of measurements reveals a rich array of properties in LaRhGe$_3$ including semimetallicity, superconductivity, and electron-phonon drag. Our electronic structure calculations further reveal topological features of interest, with the discovery of \emph{almost movable} Weyl nodal lines. \rev{This set of results on LaRhGe$_3$ highlights the important role of electron-phonon coupling and} paves the way for further exploration of the $I4mm$ class of non-centrosymmetric materials.

\section{Results and Discussion}

\subsection{Non-Centrosymmetric Structure of \LRG }\label{subsection-NonCentro}

\rev{Single crystals of \LRG\ were grown} by the metallic self-flux method~\cite{canfield1992growth}. To confirm the structural characteristics and crystalline quality of our samples, we performed powder X-ray diffraction (XRD) and energy dispersive X-ray spectroscopy (EDX). The measured powder XRD pattern, given by the blue circles in Fig.~\ref{Fig1}(a), is consistent with the previously reported body-centered tetragonal $I4mm$ (no. 107) crystal structure~\cite{venturini1985nouvelles}. Our Rietveld refinement, shown by the black line in Fig.~\ref{Fig1}(a), confirms the absence of any significant secondary phases. A minor \rev{germanium phase} (1.3\% by weight based on Rietveld refinement), likely originating from the residual flux on the surface of the crystals, \rev{is marked by the orange asterisks. No other impurity phases above the detection limit of our instrument were detected. } The fitted lattice parameters for \LRG\ are \textit{a} = 4.4187(2)~\AA\ and \textit{c} = 10.0494(5)~\AA. The full results of the Rietveld refinement are presented in Tables~\ref{RefinementSummary} and \ref{RefinementAtoms}. 

The spatial homogeneity of our crystals is confirmed by EDX mapping, which shows a uniform distribution of La, Rh and Ge on the sample surface (Fig.~\ref{Fig1}(c)). A ratio for La:Rh:Ge of 1.00:0.98:2.90 is found based on EDX analysis, which is close to the expected 1:1:3 ratio. Minor Ge-rich regions can be attributed to residual flux on the crystal's surface, as also detected in the powder XRD measurement.

The non-centrosymmetric crystal structure of \LRG\ is presented in Fig.~\ref{Fig1}(b). The broken inversion symmetry can be easily discerned by considering the positions of Rh within the unit cell, where Rh sits at the center of the $ab$-plane with the lower half of the unit cell and at the unit cell edges in the upper half of the unit cell. There are two unique Ge sites in this structure, with the Ge2 site having double the multiplicity of Ge1. The smallest interatomic distances in this structure are between Rh and Ge in a distorted corner-sharing square pyramidal configuration, where the single Rh-Ge1 distance is 2.377(3)~\AA\ and the four-fold Rh-Ge2 distance is 2.436(7)~\AA. \rev{We focus next on the implication of the tetragonal $I4mm$ space group with broken inversion symmetry on the electronic structure of \LRG .}

\begin{figure*}[t]
\centering
\includegraphics[width=18cm]{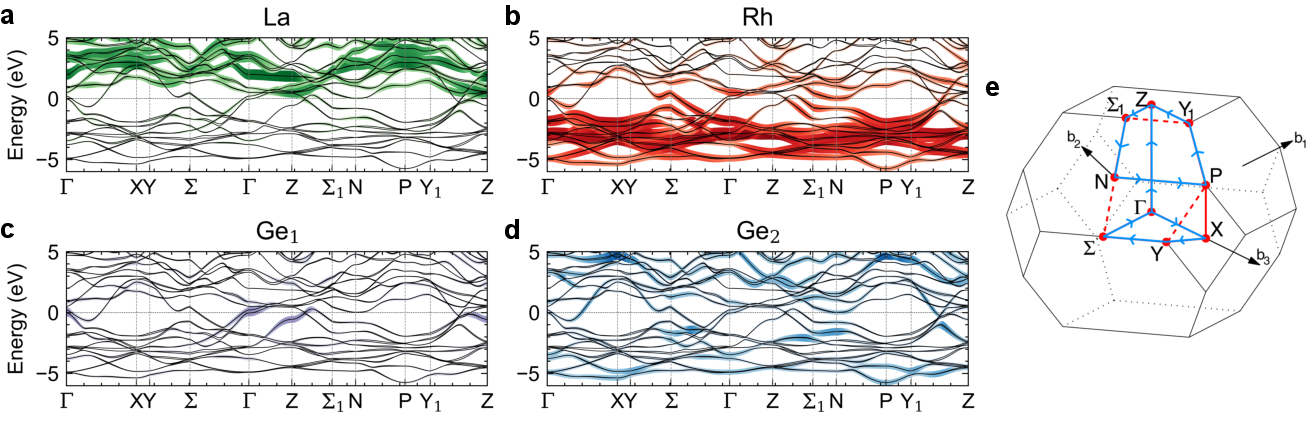}
\caption{\textit{Ab-initio} electronic band structures in the primitive unit cell of LaRhGe$_3$. The width and intensity of the superimposed colors indicate the contribution to the states from the respective atoms. \rev{The dotted horizontal black line marks the Fermi level.} An onsite Hubbard interaction of $U_{LDA+U} = 14.5$ eV was added to the $f$-orbitals of La to push the (unoccupied) states out of the energy window.}
\label{LRG-WeakCoup}
\end{figure*}

\subsection{Electronic Band Structure and Protected Nodal Lines in \LRG}\label{subsection-Nodal-Line}

We calculate the band structure of \LRG\ using density functional theory (DFT). The band structure, which is plotted along the high symmetry directions of the body-centered tetragonal Brillouin zone (Fig.~\ref{LRG-WeakCoup}), is plotted according to each atomic site (La, Rh, Ge1, and Ge2), where the thickness and intensity of the color represent the weighted atomic contribution. All four atomic sites contribute to the bands closest to the Fermi level. The largest contribution to all the bands that cross the Fermi energy originates from Rh, while Ge1 contributes to the hole bands, and Ge2 contributes to the electron bands. \rev{Due to the presence of ASOC, the spin degeneracy of all bands is lifted. The ASOC in \LRG\ was thoroughly experimentally characterized as part of a larger study on $RTX_3$ materials crystallizing in the \Ifmm\ space group~\cite{kawai2008split}. The degree of SOC on the transition element $T$ was shown to dictate the degree of splitting of spin-up and spin-down bands, which is enabled by the lack of inversion symmetry in the crystal structure. A significant splitting of the bands is realized in the case of Rh on the $T$ site~\cite{kawai2008split}, which is confirmed in our de Haas-van Alphen (dHvA) oscillations, shown in Supplemental Fig.~\ref{LRG-dHvA}. We compare our the splitting in the FFT of the dHvA signal to those previously reported~\cite{kawai2008split} in Table~\ref{dHvATable}, where we find a consistent degree of band splitting in our measurements that confirms a strong ASOC. We will discuss the implications of this strong ASOC on the other observed electronic properties of \LRG\ in the sections ahead.} 

We next highlight aspects of symmetry-enforced band crossings in \LRG. Group theoretical analysis recently predicted \textit{almost movable nodal lines} in a number of tetragonal space groups including the \Ifmm~group to which \LRG\ belongs~\cite{hirschmann2021symmetry}. Their specific dispersion relation, connectivity, and proximity to the Fermi level are not fully fixed by symmetry alone and are instead found to be material-dependent. 
In the case of \LRG, even though the global Kramers' degeneracy is lifted, the $I4mm$ space group symmetry enforces two types of two-fold degenerate nodal lines throughout the Brillouin zone. Consider the mirror plane that intersects the rotational axis ($\Gamma$ - $Z$) and the $N$-point. The eigenvalues of the mirror symmetry at $N$ are $\pm i$. Because $N$ is a time-reversal invariant momentum (TRIM) point, the $+i$ and $-i$ eigenvalues are paired, enforcing a two-fold degeneracy at this point. Let $\mathbf{k}_1$ and $\mathbf{k}_2$ be two $k$-points on the mirror plane that are related by time reversal. Because they are time reversal partners, their mirror eigenvalues have opposite phases. Therefore, along an arbitrary path on the mirror plane that connects $\mathbf{k}_1$ and $\mathbf{k}_2$ (which does not cross a TRIM point), the eigenvalues must be exchanged. 
Since this can only happen at band crossings and we have the freedom to choose any path, the intersections must form a nodal line that connects the $N$-points. That type of degeneracy is called an \textit{almost movable} nodal line~\cite{hirschmann2021symmetry}. It is only pinned at the two TRIMs that it connects and, in contrast to a point-like topological band feature (\emph{e.g.} a Weyl point), it is extended in both $k$-space and energy. The same argument holds for the $\Gamma$- and $Z$-point. However, the presence of additional symmetry pins the nodal line to the rotational axis. Hence, there are two ways the almost movable nodal line can connect the $N$-points: It either intersects the pinned nodal line ($\Gamma$ - $Z$) or it does not.

\begin{figure}
    \centering
    \includegraphics[width=8.6cm]{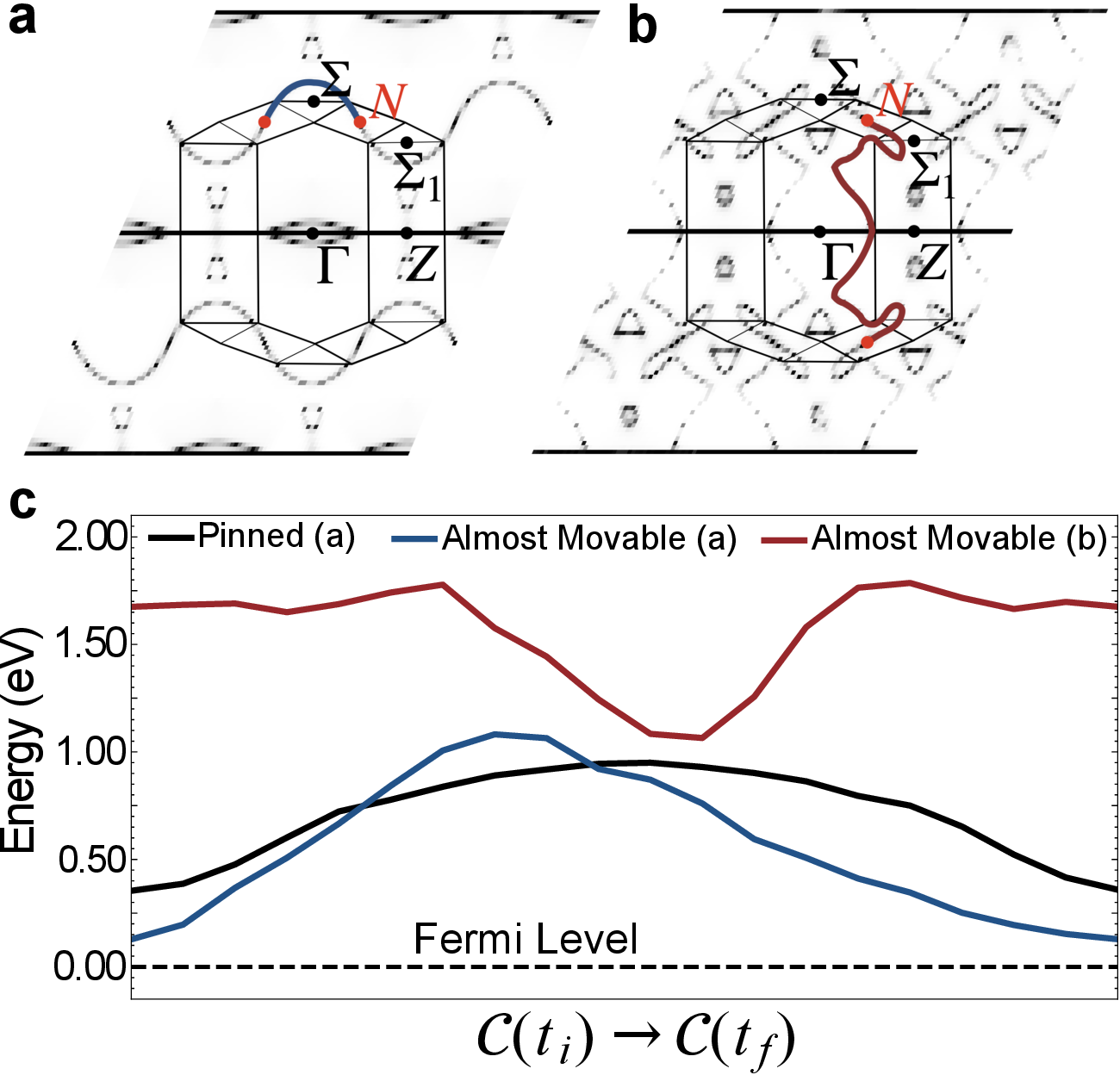}
    \caption{Almost movable nodal lines in LaRhGe$_3$. (a,b) Pinned ($\Gamma$ to $Z$, black) and almost movable nodal lines ($N$ to $N$, (a) blue and (b) red) in the mirror plane of the Brillouin zone of the type 2 body centered tetragonal primitive unit cell. The projection of the Brillouin zone onto this mirror plane, as well as the high-symmetry points that lie therein are given in the overlay. The connectivity of almost movable nodal lines is not completely determined by the space group symmetry. The two possibilities are both realized in the material for different pairs of bands. (a) The almost movable nodal line connecting two $N$-points. (b) Same as (a), but intersecting the pinned nodal line. (c) The dispersion along curves $\mathcal{C}$ through $k$-space along the pinned nodal line from (a) (black), the almost movable nodal line from (a) (blue), and the almost movable nodal line from (b) (red). $t$ parameterizes the curve with $\mathcal{C}(t_i)$ and $\mathcal{C}(t_f)$ being the initial and final point of the curves respectively. The pinned and the almost movable lines from (a) are close to the Fermi level.}
    \label{fig:almost_movables}
\end{figure}

In Figure~\ref{fig:almost_movables}(a) and \ref{fig:almost_movables}(b) the two types of connectivity are shown. The intensity of the heatmap corresponds to a quantity similar to a spectral function that has poles at $k$-points where a chosen pair of bands becomes degenerate. The overlay shows the body-centered primitive Brillouin zone (type 2) and the high-symmetry points that lie within the plane spanned by $\Gamma$-$Z$ and $\Gamma$-$N$. The almost movable nodal lines connect the $N$-points on (seemingly arbitrary) paths that depend on the specific dispersion relation of the chosen pair of bands. Thus, the evolution of the almost movable nodal lines through $k$-space for different band pairs varies drastically. 

Figure~\ref{fig:almost_movables}(c) shows the dispersion relation along the different nodal lines. The pinned and the almost movable nodal lines shown in Fig.~\ref{fig:almost_movables}(a) are close to the Fermi energy and can potentially be tuned to the Fermi edge by either doping or applying a gate voltage. Due to the dispersive nature of the nodal lines, there is a large energy region in which the degeneracies can be accessed in contrast to point-like topological band crossing, and thus a wider range of parameters in which drumhead surface states can be found~\cite{weng2015topological,chiuPRB2016nodalline}. Even though surface states do not necessarily emerge in nodal line systems~\cite{chiu_schnyder_arXiv_18}, any curve through $k$-space that encircles a single nodal line has a Berry curvature that is quantized to $\pi$ by the mirror symmetry. A suitable surface termination can likely be found for which gapless surface states appear at $k$-points of the two-dimensional surface Brillouin zone that connect projections of the different nodal lines in the bulk. 

Materials with these gapless and robust surface modes are highly sought-after, due to the variety of proposals in novel device applications~\cite{liu2019topological, burkov2016topological, tian2015electrical, tian2017observation, li2018spin, kim2015observation, lundgren2014thermoelectric, wang2017ultrafast}. In realistic materials, disorder and defects weakly break the crystal symmetries that protect both the nodal lines and their edge modes. However, as long as the symmetries are approximate and the effect of the perturbation is small compared to the topological gap, the edge modes remain present~\cite{Guo2022,noncrystallinequasisymmetry}. Interaction effects might also weakly break symmetries, but more importantly they generically obstruct an exact description of a material's electronic structure using single-particle states, which the definition of most topological invariants relies on. Nevertheless, it has been shown that an interacting system can always be adiabatically connected to a noninteracting system with the same topology, as long as the interaction does not drive a phase transition~\cite{topHam3, topHam1, topHam2, topHam4}. Even nonlinear perturbations do not generically suppress edge modes~\cite{stability_nonlinear}, and their stability in nodal line semimetals in their superconducting state has been theoretically demonstrated~\cite{stableSC}. \rev{Therefore, realizing superconducting state, even a conventional one mediated by electron-phonon coupling, in the bulk of a nodal line semimetal can allow for exotic pairing on the surface due to interaction with the topological surface state. }

\begin{figure*}[tbh]
\centering
\includegraphics[width=18cm]{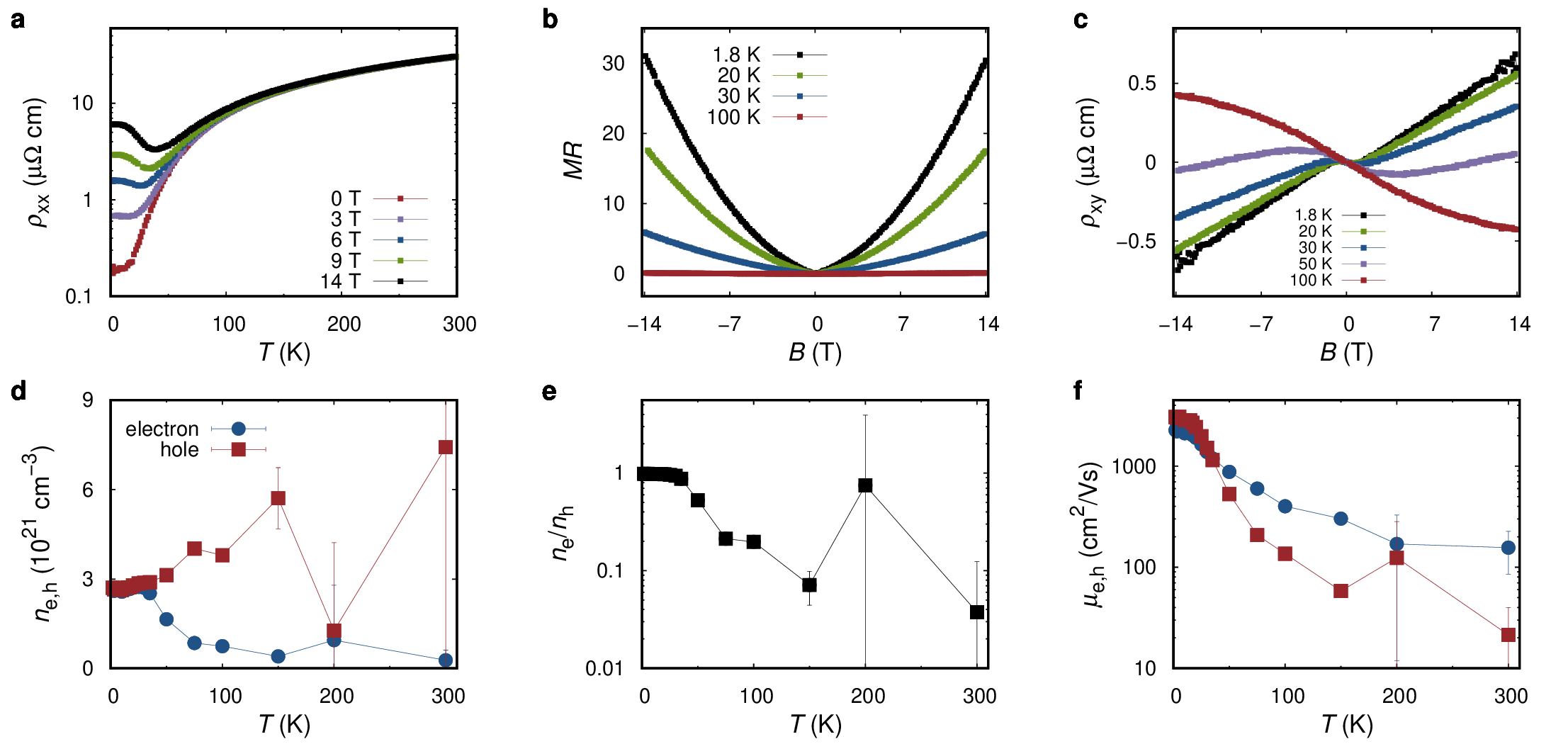}
\caption{Semimetallic electronic properties of LaRhGe$_3$. (a) Temperature dependence of the longitudinal resistivity $\rho_\mathrm{xx}$ measured in different magnetic fields, showing sizable magnetoresistance below 50 K. (b) Magnetic field dependence of the resistance (MR) as defined in text measured at various temperatures, , reaching a value of 30 by 14\,T at 1.8\,K. (c) Hall resistivity $\rho_\mathrm{xx}$ measured at various temperatures, showing a change of sign upon cooling below 50\,K. (d) Electron and hole carrier densities, (e) their ratio, and (f) carrier mobilities extracted from fits of MR and $\rho_\mathrm{xy}$ to a two-band model~\cite{pippard1989magnetoresistance}, revealing LaRhGe$_3$ to be a compensated semimetal at low temperatures.}
\label{Fig2}
\end{figure*}

\subsection{Semimetallicity in \LRG }\label{subsection-Semimetal}

\rev{We next highlight the semimetallic nature of \LRG\ experimentally, which, along with our nodal-line predictions, demonstrate that it is a Weyl semimetal.} Electrical transport measurements confirm the metallic nature of this material, with a longitudinal resistivity that monotonically decreases with decreasing temperature and resistivity magnitudes in the $\mu\Omega$-cm range across all temperatures (Fig.~\ref{Fig2}(a)). The high residual resistivity ratio, $\rho_\mathrm{300\rm{K}}/\rho_\mathrm{2\rm{K}} = 110$, is suggestive of the good crystalline quality of our samples. As a function of increasing applied magnetic field, there is a rapid increase in the low temperature resistivity, below 50\,K. For fields larger than 6~T, the resistivity reaches a minimum value before a low temperature upturn and then ultimately plateaus for temperatures below $10~\rm{K}$. In Fig.~\ref{Fig2}(b), we plot the magnetoresistance (MR) of \LRG\ at various temperatures, where MR is defined as $[\rho_\mathrm{xx}(B)-\rho_\mathrm{xx}(0)]/\rho_\mathrm{xx}(0)$. The MR remains unsaturated in fields up to 14\,T, where it reaches a value of $30$ at 1.8\,K. The large and unsaturated MR and the field-driven resistivity upturn in \LRG\ are characteristic of a compensated semimetal~\cite{wang2016resistivity,wang2017large,bannies2021extremely}. \rev {We perform a scaling analysis of the MR over the various temperatures measured, as shown in the supplemental materials Fig.~\ref{LRG-Kohler}, and find good agreement with Kohler's rule.} 

\rev{To verify the semimetallicity scenario, we measured the Hall resistivity $\rho_\mathrm{xy}$ of LaRhGe$_3$, shown in Fig.~\ref{Fig2}(c), at the same fields and temperatures as the MR. We find that $\rho_\mathrm{xy}(B)$ is non-linear for all temperatures, and hence it cannot be fitted with a semi-classical one-band model for which $\rho_\mathrm{xx}$ would be independent of field, also in disagreement with our data. 
This is expected if we examine the band structure of \LRG , shown in Fig.~\ref{LRG-WeakCoup}, where we find three pairs of spin-split bands crossing the Fermi level, where two pairs are hole bands and one pair is an electron band. While the electron Fermi surfaces are larger than the hole Fermi surfaces, we fitted both the MR and Hall resistivity to a semi-classical two-band model~\cite{pippard1989magnetoresistance} that assumes density and mobility of all hole and electron carriers can each be captured with a single term:}

\begin{equation}
    \rho_\mathrm{xx}=\frac{1}{e}\frac{(n_\mathrm{h}\mu_\mathrm{h}+n_\mathrm{e}\mu_\mathrm{e})+(n_\mathrm{h}\mu_\mathrm{e}+n_\mathrm{e}\mu_\mathrm{h})\mu_\mathrm{h}\mu_\mathrm{e}B^2}{(n_\mathrm{h}\mu_\mathrm{h}+n_\mathrm{e}\mu_\mathrm{e})^2+(n_\mathrm{h}-n_\mathrm{e})^2\mu_\mathrm{h}^2\mu_\mathrm{e}^2B^2}
\end{equation}

\begin{equation}
    \rho_\mathrm{xy}=\frac{B}{e}\frac{n_\mathrm{h}\mu_\mathrm{h}^2-n_\mathrm{e}\mu_\mathrm{e}^2+(n_\mathrm{h}-n_\mathrm{e})\mu_\mathrm{h}^2\mu_\mathrm{e}^2B^2}{(n_\mathrm{h}\mu_\mathrm{h}+n_\mathrm{e}\mu_\mathrm{e})^2+(n_\mathrm{h}-n_\mathrm{e})^2\mu_\mathrm{h}^2\mu_\mathrm{e}^2B^2}.
\end{equation}
The fitting parameters in this model are the carrier densities, $n_\mathrm{e}$ and $n_\mathrm{h}$, and the carrier mobilities, $\mu_\mathrm{e}$ and $\mu_\mathrm{h}$, for electrons and holes, respectively. We simultaneously fit $\rho_\mathrm{xx}$ and $\rho_\mathrm{xy}$ to this two-band model to extract $n_\mathrm{e,h}$ and $\mu_\mathrm{e,h}$ as a function of temperature. \rev{The carrier densities, plotted in Fig.~\ref{Fig2}(d), are on the order of $10^{21}$~cm$^{-3}$, a high carrier density semimetal, such as HfB$_2$~\cite{wang2019magnetotransport} and LaAlGe~\cite{bhattarai2021experimental} with similar carrier densities.} 

At high temperatures, we find that holes are the majority charge carriers but when the temperature falls below 40\,K, the carrier densities are essentially compensated with $n_\mathrm{e}/n_\mathrm{h}=0.99(1)$ at 2\,K, as can be seen in Fig.~\ref{Fig2}(e). For both holes and electrons, the mobility increases with decreasing temperature, as can be expected due to reduced scattering of charge carriers by phonons. \rev{At high temperature the electron mobility is larger than the hole mobility, as can be seen in Fig.~\ref{Fig2}(f). We should note, however, that the error bars in this fitting range are large due to the small magnetoresistance above 100 K and the limitations of the two band model.} Upon cooling, the hole mobility increases more rapidly and exceeds the electron mobility for temperatures where the carriers are compensated, below $40$\,K, explaining the sign change observed in $\rho_{xy}$ in the same temperature range. The high hole mobility of 0.3~m$^2$/Vs at low temperature again demonstrates the good crystal quality, which is also relevant to the non-saturating MR of \LRG\ \rev{and the electron-phonon interactions discussed in Sec.~\ref{subsection-e-ph-drag}.
We have experimentally verified the semimetallic nature of \LRG , and in the next section demonstrate the discovery of superconductivity in this Weyl semimetal.}

\subsection{Superconductivity in \LRG }\label{subsection-SC}

\begin{figure*}[t]
\centering
\includegraphics[width=18cm]{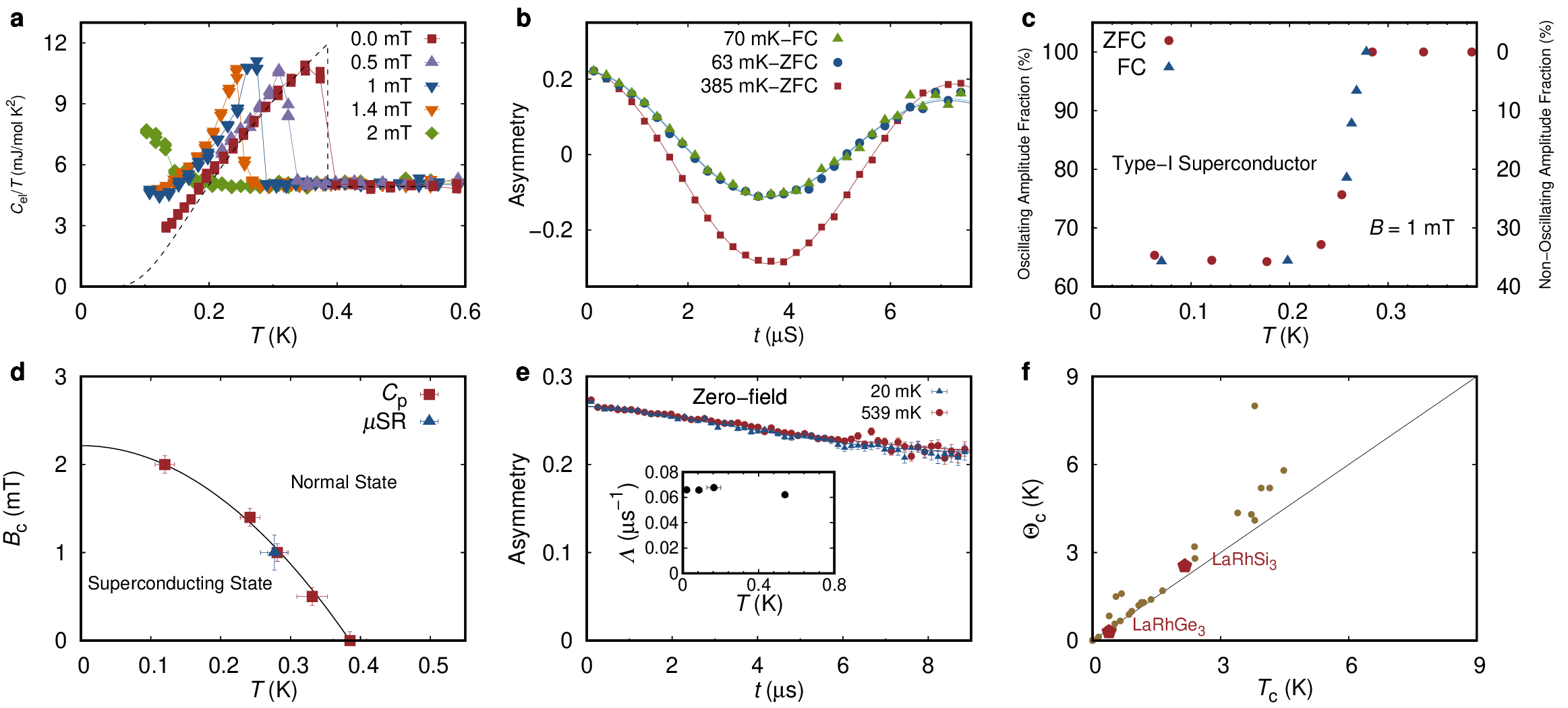}
\caption{Superconducting transition in \LRG. (a) Temperature dependence of \Ce /$T$ measured in varying magnetic fields. The superconducting transition seen at 0.39(1)\,K in 0~mT is suppressed with increasing magnetic field. (b) Representative transverse field muon decay asymmetry spectra collected above and below \Tc\ in \LRG\ under an applied magnetic field of 1.0~mT. \rev{(c) The oscillating and non-oscillating amplitude of the signal in (b) as a function of temperature measured in zero-field cooling (ZFC) and field cooling (FC) protocol. Reduction in the oscillating amplitude is due to expulsion of magnetic field in the superconducting state, and the ZFC and FC show similar suppression, demonstrating type-I superconductivity. (d) Plot of $\Theta_c$ against measured \Tc\ for various elemental type-I superconductors, LaRhSi$_3$, and \LRG . \LRG\ falls on the black solid line, which represents the expectation based on weak-coupling BCS theory. (e) Field dependence of the \Tc\ values extracted from \Cp\ and transverse field \muSR. The solid line is a fit to the expression, Eq. 7, in the text to estimate \mHc~$ = 2.2$~mT. (f) Zero field \muSR\ spectra collected at 539~mK and 20~mK, with fit using Kubo-Toyabe function for the data. No significant change is seen between the data above and below \Tc . Inset shows the resulting relaxation rate $\sigma$.}}
\label{LRG-sc}
\end{figure*}

Our first indication of superconductivity in \LRG\ comes from low temperature specific heat measurements\rev{. We observe a sharp lambda-like anomaly at \Tc\ $= 0.39(1)$\,K as can be seen in the electronic part of the specific heat \Ce , as shown in Figure~\ref{LRG-sc}(a), where the lattice contribution has been subtracted by fitting the zero-field data above the transition.} Upon application of small magnetic fields, the transition is suppressed to lower temperatures. We find that the transition remains sharp for applied fields of 0.5~mT and 1~mT, despite the decrease in \Tc, \rev{which is consistent with the first order field-induced suppression of superconductivity expected for a type-I superconductor~\cite{svanidze2012type,zhao2012type,sun2016type} and comparable to what has previously been observed in LaRhSi$_3$~\cite{anand2011specific}.}
Upon application of a 2~mT field, the superconducting transition is fully suppressed in our measurement window, which extends down to 0.1~K. The magnitude of the specific heat jump, $\Delta C/C_{en} \sim 1.43$, as shown in shown in Fig.~\ref{LRG-type-I}(a), is consistent with a BCS model in the weak-coupling limit.
\Cp\ data before subtraction of the phonon part are shown in Fig.~\ref{LRG-Cp}.
\rev{Further analysis of this specific heat data, as described in the Supplemental Materials, allows a number of superconducting parameters to be derived, and these are presented in Table~\ref{SCTable}.}

%

We next present \rev{muon spin relaxation and rotation ($\mu$SR) measurements on \LRG. Muons are highly sensitive to their local magnetic environment, making them an excellent probe of field expulsion in superconductors and spontaneous time-reversal symmetry breaking. We begin by investigating the former, where} the muon decay asymmetry plotted as a function of time in an applied field of 1~mT (which is below the critical field) is shown in Fig.~\ref{LRG-sc}(b). \rev{In this geometry, the direction of the magnetic field is transverse to the muon spin. Above the superconducting transition, at 385 mK, the muons that land in the sample or the sample holder will couple to the applied field and undergo Larmor precession, yielding an oscillatory decay asymmetry. Whereas, within the superconducting state at 70~mK, the magnetic field is expelled and the muons that land in the sample will not precess, giving rise to a flat decay asymmetry. The remnant oscillatory signal originates from the fraction of muons that land outside the superconducting sample.} 
We fit the asymmetry spectra using a two-term asymmetry function
\begin{equation}
\begin{split}
G_{\mathrm{TF}}(t) & =A\left[f \exp \left(-\lambda t \right) \cos \left(\omega_{1} t+\phi\right)\right. \\
& \left.+(1-f) \exp \left(\frac{-\sigma^{2} t^{2}}{2}\right)\right]
\end{split}
    \label{muSR_TF_fit}
\end{equation}
\noindent
where the first term captures the signal from muons stopping in the silver sample holder and non-superconducting part of the sample and the second term captures the signal from muons stopping in the superconducting part of the sample. The fraction of signal from muons that land in silver sample holder and non-superconducting part of the sample is given by $f$, and $\omega_{1}$ is the muon precession frequencies in the background. The $A$ term is the total asymmetry and the $\phi$ is the initial phase of the muons. The depolarization rates for the sample and the background signals, respectively, are given by $\sigma$ and $\lambda$. A global fit was used for all variables, but the fraction $f$ was allowed to be refined with temperature. The value of $f$ as a function of temperature under both ZFC and FC conditions is shown in Fig.~\ref{LRG-sc}(c). The $100\%$ oscillating amplitude corresponds to all muons stopping inside \LRG\ above \Tc\ and the non-superconducting silver plate. A sharp drop in the oscillating amplitude is observed at the superconducting transition. This behavior is expected in a superconductor due to the expulsion of the magnetic field from the bulk of the sample in the superconducting state such that muons implanted sufficiently deeply in \rev{the sample will not be affected by the applied field and will therefore not precess. We estimate about 40\% of the signal amplitude corresponds to the sample based on the sample coverage, so the drop to about 60\% corresponds to the expulsion of the field from most of the sample. We show a shift in the amplitude from the oscillating component to the zero-field in Fig.~\ref{LRG-muSR-FFT}.} From these transverse field $\mu$SR measurements, we estimate \Tc $\sim0.28$\,K in an applied field of 1~mT, in good agreement with the \Tc\ measured in specific heat at 1~mT. We find that the suppression of the amplitude is approximately equivalent under ZFC and FC conditions at the lowest temperatures, which further supports the conclusion that \LRG\ is a Type-I superconductor.



\begin{table}[htbp]
\centering
\caption{Superconducting parameters of \LRG.}
\label{SCTable}
\setlength{\extrarowheight}{3pt}
\begin{tabular}{>{\centering}m{1.7cm} >{\centering}m{2.3cm} >{\centering\arraybackslash}p{2.5cm}}
\hline
\hline
\textbf{Parameter}     & \textbf{Value} & \textbf{Units}   \\
\hline
\Tc                          & 0.39(1)  & K  \\
\mHc                   & 2.2(1)  & mT    \\
$\gamma_n$            & 4.92(4)  & mJ mol$^{-1}$ K$^{-2}$    \\
$\beta$           & 0.25(1)   & mJ mol$^{-1}$ K$^{-4}$     \\
$\theta_D$                & 341(10)   & K          \\
$\lambda_{ep}$                   & 0.34(2)    & -      \\
\hline
\hline
\end{tabular}
\end{table}

The temperature and field dependence of \Tc\ extracted from specific heat and transverse field $\mu$SR are plotted in Fig.~\ref{LRG-sc}(d). We estimate the zero-temperature critical field \mHc\ using the conventional relationship, $B_{c}(T) = B_{c}(0) \left[1- \left(\sfrac{T}{T_c}\right)^2 \right]\ $,
where \Tc\ is $0.39(1)$\,K. This fit yields the curve shown by the black line and a \mHc\ of 2.2~mT. 
The small critical field value makes the presence of a remnant magnetic field during the \Cp\ measurement a serious concern that can shift our measured \Tc . However, to address this concern we highlight the good agreement with the point on this phase diagram extracted from transverse field \muSR\ measurement, shown in blue. This point is expected to be accurate due to the high-resolution field-zeroing protocol employed in that experiment. 

\rev{Our attempts to measure the electric resistivity of \LRG\ in a dilution refrigerator reveal a drop in the resistivity at a temperature that is commensurate with the onset of superconductivity, as shown in Fig.~\ref{LRG-Cp}(b). This resistance drop is partially suppressed by an applied magnetic field of 1.0~mT. However, even in the zero field measurement, we detect a finite resistance in the superconducting state, which we ascribe to the applied current exceeding the critical current threshold of \LRG . The low critical current suggested by our resistivity measurements is not particularly surprising given the extremely low critical field of 2.2~mT.} 


Finally, we performed zero field $\mu$SR measurements to look for any sign of spontaneous magnetic fields and time-reversal symmetry breaking as \LRG\ enters its superconducting state. The magnetic field at the sample position was carefully zeroed following the protocol described in the methods. High statistics zero field spectra were collected above and below $T_c$ with representative spectra shown in Fig.~\ref{LRG-sc}(e). In the absence of magnetism, the muon depolarization is due to randomly oriented nuclear moments and can be described by the Kubo-Toyabe function
$$
G_{\mathrm{KT}}(t)=\frac{1}{3}+\frac{2}{3}\left(1-\sigma^{2} t^{2}\right) \exp \left(-\frac{\sigma^{2} t^{2}}{2}\right)
$$
\noindent
where $\sigma$ accounts for the field generated by the randomly nuclear moments at the muon site. The relaxation spectra for \LRG\ is fitted to the function
$$
A(t)=A_{S} G_{\mathrm{KT}}(t) \exp (-\Lambda t)+A_{\mathrm{BG}}
$$
\noindent
where $A_S$ and $A_{BG}$ represent the sample and background asymmetry, respectively, while $\Lambda$ accounts for any additional relaxation rate. Our fits reveal no significant change in $\Lambda$ above and below \Tc , as shown in the inset of Fig.~\ref{LRG-sc}(e), confirming the absence of any significant TRS breaking in \LRG .

Type-I superconductivity is common among pure elements but relatively rare in compounds, particularly beyond binaries. 
\LRG\ is therefore one of very few~\cite{hull1981superconductivity,palstra1986superconductivity,yonezawa2005type,anand2011specific} type-I superconductors containing more than two elements.
In Fig.~\ref{LRG-sc}(f), we compare the calculated critical temperature, denoted as $\theta_c$, based on the $\gamma_n$ and \mHc\ values
~\cite{marques2020electronic}, with the measured \Tc\ for various elemental type-I superconductors, $\rm{La}\rm{Rh}\rm{Si}_3$~\cite{anand2011specific}, and \LRG . We find that \LRG , as well as $\rm{La}\rm{Rh}\rm{Si}_3$, lie close to the solid-black line that represents the expectation based on weak-coupling BCS, further confirming that \LRG\ is a type-I superconductor in the weak-coupling limit. 
\rev{Nevertheless, we expect topologically protected states on the surface of \LRG , as described in Section~\ref{subsection-Nodal-Line} may allow for novel pairing on the surface despite the conventional nature of the bulk. Next we highlight electron-phonon interactions in the normal state of \LRG , which, unlike the superconducting state, hint at strong coupling.}

\begin{figure*}[t]
\centering
\includegraphics[width=12.5cm]{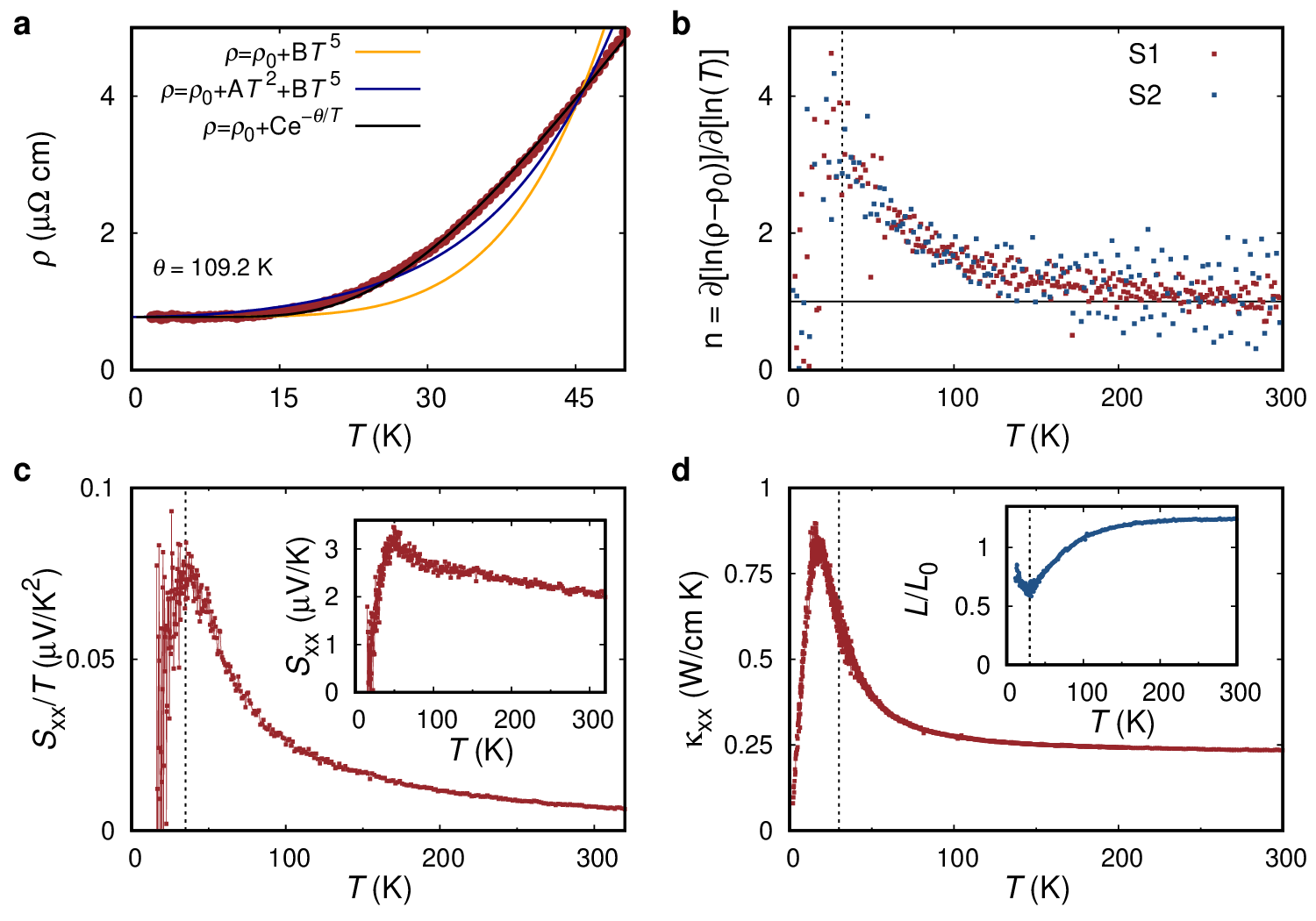}
\caption{Phonon drag in LaRhGe$_3$ detected by transport measurements. (a) Resistivity as a function of temperature measured along $ab$-plane of \LRG\ in zero-field, and different models fitted to the data below 50\,K. (b) $n$,  the exponent of the temperature dependent resistivity determined derivative of $ln(\rho-\rho_0)$ vs. $ln(T)$ as a function of temperature. Data for two samples, S1 and S2, are shown. (c) Temperature dependence of Seebeck coefficient over temperature \rev{($S_{xx}/T$)} as a function of temperature of \LRG . Inset shows $S_{xx}$ as a function of temperature. (d) Temperature dependence of thermal conductivity of \LRG . Inset shows the Lorenz number as a function of temperature. The dashed vertical line in panels (b,c,d) marks the approximate phonon drag temperature.}
\label{LRG-ph}
\end{figure*}

\subsection{Electron-Phonon Drag in \LRG }\label{subsection-e-ph-drag}

Finally, we return to the normal state of LaRhGe$_3$ to investigate the thermal properties of this material. Our first indication of unusual transport behavior comes from zero-field resistivity where we observe a deviation from a standard Fermi liquid behavior as shown in Fig.~\ref{LRG-ph}(a). Even using a modified Bloch-Grüneisen model $\rho=\rho_0+AT^{2}+BT^{5}$, where the $T^{5}$ term accounts for electron-phonon interactions and the additional $T^2$ term accounts for electron-electron interactions, a poor fit is obtained below 50 K. 
\rev{Instead, we find that the resistivity data are better described by the exponential form of a phonon-drag model, $\rho=\rho_0+Ce^{-T_0/T}$~\cite{hicks2012quantum,yang2021evidence}, where the activation temperature is $T_0=110$\,K. 
We find an equally good agreement with the data using a \rev{power law fit $\rho=\rho_0+CT^{n}$,} which yields $n=2.9$, as shown in Sup.~Fig.~\ref{LRG-Tn}.} 
To further analyze the temperature dependence of the resistivity we take the logarithmic derivative after subtracting residual resistivity to quantify the exponent $n={[\delta ln(\rho-\rho_0)]/[\delta lnT]}$. 
The temperature dependence of $n$ is presented in Fig.~\ref{LRG-ph}(b) for two different samples, and we find at high temperature $n\sim1$ as expected in a standard Block-Gruneisen picture. However, we find an upturn towards $n\sim3$ at low temperatures, which is consistent with $n=2.9$ from our power law fit. 

\begin{figure*}[t]
\centering
\includegraphics[width=18.0cm]{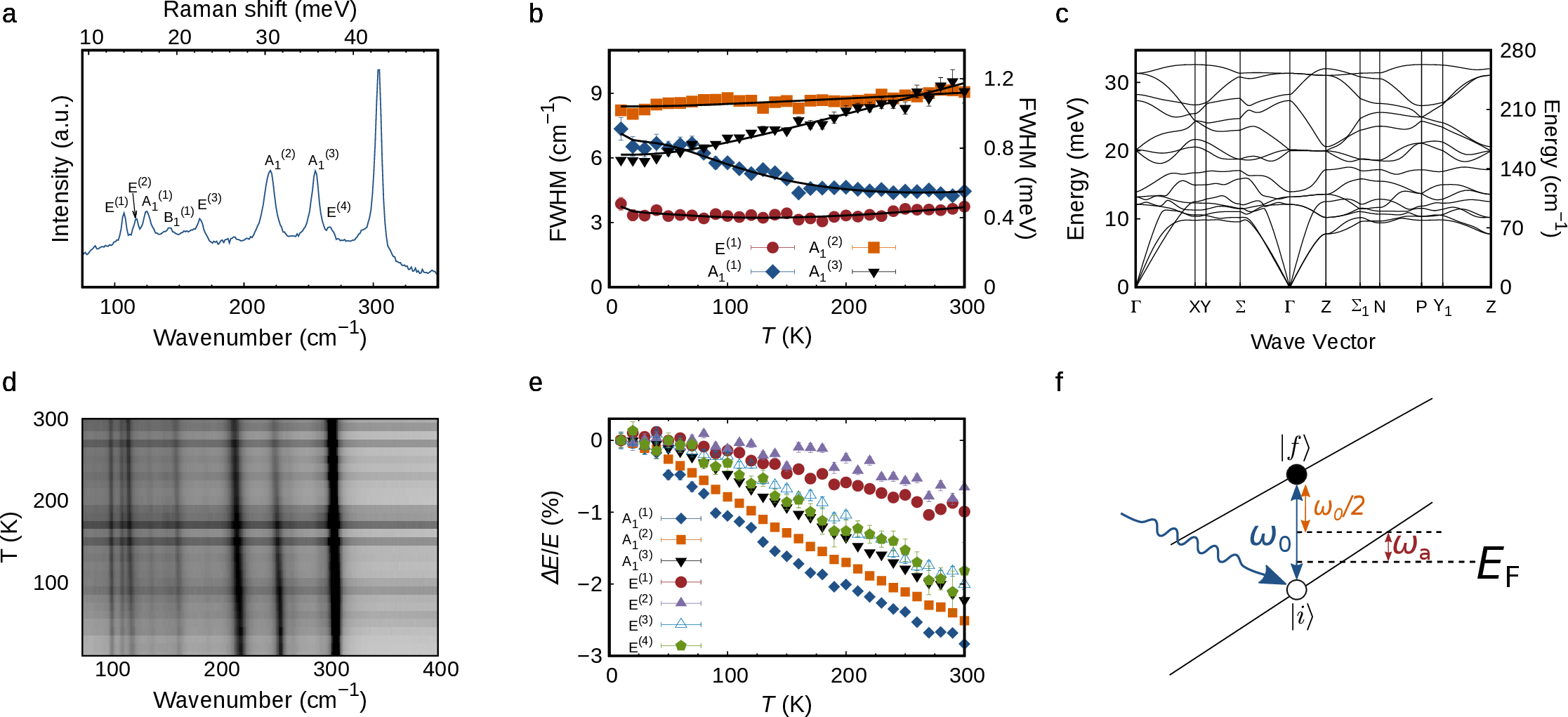}
\caption{\rev{Electron-phonon drag in LaRhGe$_3$ detected by Raman spectroscopy. (a) Raman spectra of \LRG\ measured at 10~K. The phonon mode symmetries are labelled in accordance with our DFT calculations. The strong peak at 300~$\rm{cm}^{-1}$ is due to Ge flux on the surface. (b) Black and white map of Raman scattering intensity plotted as a function of temperature and Raman shift measured from the (101) surface of \LRG\ crystal. (c) Full width at half maximum (FWHM) temperature dependence of $A_1^{(1)}$, $A_1^{(2)}$, $A_1^{(3)}$, and $E^{(1)}$ modes based on fits described in the text. Solid black lines correspond to fits to the temperature dependence of each phonon mode based on Eq.~\ref{ahmW}.  (d) Percent difference of the phonon energies of $A_1$ and $E$ modes from energies at 10~K as a function of temperature. (e)  Calculated phonon dispersion curves along high-symmetry lines of the tetragonal primitive unit cell of \LRG . Anomalous temperature dependence of the $A_1^{(1)}$ phonon lifetime in LaRhGe$_3$. (f) Schematic depicting excitation of electron-hole pairs by optical phonons, where $\omega_a$ is the energy offest to take into account that the Fermi level is not exactly halfway between the initial state $\left |  i \right \rangle$ and final state $\left |  f \right \rangle$ and $\omega_0$ is the bare phonon energy.}}
\label{LRG-Raman}
\end{figure*}

Further evidence for the phonon-drag scenario comes from Seebeck coefficient measurements, shown in the inset of Fig.~\ref{LRG-ph}(c), where we notice a maximum at around $\sim50$\,K signaling the interaction between phonons and electrons is enhanced at low temperature. \rev{When scaled by temperature, $S_{xx}/T$, plotted in the main panel of Fig.~\ref{LRG-ph}(c), the peak appears at a temperature of $\sim30$~K similar to the peak in the derivative of the resistivity.} Coincident with these two features, a peak also appears below 50~K in the thermal conductivity, shown in Fig.~\ref{LRG-ph}(d). Typically, the Lorenz number is evaluated by dividing the electronic part of the thermal conductivity by the electrical conductivity, but separating the phonon and electron parts of the thermal conductivity proved difficult with our current data.
Instead, we show the total thermal conductivity divided by electrical conductivity, representing an upper limit of the Lorenz number, in the inset of Fig.~\ref{LRG-ph}(d), which shows clear deviation at low temperature with a minimum around 30~K. It's worth noting that no clear deviations in the scaling behavior of the MR is detected at this temperature, as shown in Fig.~\ref{LRG-Kohler}.

The peaks in $\kappa_{\rm{xx}}$, $S_{\rm{xx}}/T$, and the resistivity derivative all appear close to $\sim30$~K, which we take as the phonon-drag temperature. Comparing this with the Debye temperature or 340~K, we find $\Theta_{\rm{ph-drag}}/\Theta_{\rm{D}}\sim10$, which has been seen in other materials with phonon-drag~\cite{}. \rev{In Sec.~\ref{subsection-Semimetal}, we noted that the MR is significantly enhanced below 30\,K due to electron-hole compensation. At the same temperature, we also observed a crossover of the carrier mobilities with the holes having higher mobilities in the low-temperature limit and both electron and hole mobilities plateauing in this temperature range. Therefore, the low-temperature increase in $\kappa_{\rm{xx}}$ and $S_{\rm{xx}}/T$ can be attributed to 
and electron-phonon interactions dominating over phonon-phonon interactions.} 

To confirm this scenario, we measured the Raman spectra of LaRhGe$_3$ to search for signatures of phonon linewidth broadening due to electron-phonon coupling. 
Due to the lack of inversion symmetry, all 12 optical phonons in \LRG\ are Raman active, which we will denote by the irreducible representations of the $C_{4v}$ point group to \rev{8 optical phonons, where 4 of them are doubly degenerate}: $3A_1+ B_1+ 4E$~\cite{BilbaoRaman}.
Using the Raman selection rules and polarization optics~\cite{Loudon}, we identify the symmetry of each peaks at 121, 214, and 249\,cm$^{-1}$ as the three $A_1$ phonons, the peak at 139\,cm$^{-1}$ is a $B_1$ phonon, whereas the peaks at 106, 116, 163, and 262\,cm$^{-1}$ are the phonons with $E$ symmetry (Fig.~\ref{LRG-RamanPolarization}). 
The Raman spectrum measured at 10\,K without any analyzing polarizer is shown in Fig.~\ref{LRG-Raman}(a) with all phonon modes labeled. 
We tabulate the measured phonon energies alongside the values calculated from DFT in Table~\ref{RamanTable} to justify our mode assignment.
We also notice that there are two unaccounted peaks at 153 and 189\,cm$^{-1}$, which we designate as phonon overtones, \emph{i.e.} the simultaneous excitation of two phonons carrying opposite momenta.
Our assignment is justified by the DFT calculated phonon  (Fig.~\ref{LRG-Raman}(e)), where the acoustic phonon branch has maximum energy around 80-90\,cm$^{-1}$.

The Raman spectra were measured between 10 and 300\,K in 10\,K intervals \rev{as shown in Fig.~\ref{LRG-Raman}(b) and Fig.~\ref{LRG-Raman-Sup}.} 
In order to maximize the signal-to-noise ratio and track all phonon modes simultaneously, the temperature-dependent measurements were done without the analyzing polarizer.
The phonon energies and full-width at half maxima (FWHM) were extracted by fitting to Lorentzian line shapes, as shown in Fig.~\ref{LRG-Raman-Fit-Sup} for spectra collected at 10~K and 300~K. \rev{The resulting temperature dependence of the FWHM and energy of select phonon modes are presented in Fig.~\ref{LRG-Raman}(c) and (d), respectively.
At a qualitative level, we note that the energy shifts show a conventional shift to lower energies with increasing temperature for all modes. Meanwhile, a more complex behavior is uncovered for the FWHM, with some modes showing an unexpected narrowing with increasing temperature.}

In most materials, the phonons acquire finite linewidths through three channels: impurity scattering ($\gamma_0$), anharmonic phonon-phonon interaction ($\gamma_{3p}(T)$), and electron-phonon coupling ($\gamma_{ep}(T)$)~\cite{Bonini2007PRL,PBAllen}:
\begin{equation}\label{ahmW}
\begin{multlined}
    \gamma(T) = \gamma_0 + \gamma_{3p}[1+n(\omega_1,T)+n(\omega_0-\omega_1,T)] \\
    + \gamma_{ep}[f(\omega_a-\omega_0/2)-f(\omega_a+\omega_0/2)].
\end{multlined}
\end{equation}
The temperature dependence of the FWHM is typically dominated by the anharmonic ``three-phonon process'' given by the second term in Eq.~\ref{ahmW}. Here, a $q=0$ optical phonon of $\omega_0$ decays into two phonons with opposite momenta and energies $\omega_1$ and $\omega_0-\omega_1$, and $n(\omega,T)$ 
is the Bose distribution function.
The last term in Eq.~\ref{ahmW} accounts for the line width due to the excitation of an electron from state $(k,i)$ to $(k,j)$ by absorbing an optical phonon of energy $\omega_0$ (Fig.~\ref{LRG-Raman}(f)), where $f(\omega,T)$ 
is the Fermi distribution function.
The constant $\gamma_{ep}$ is directly related to the electron-phonon coupling constant, $\lambda=\frac{4}{\pi N(0)}\sum_{j} \gamma_{ep,j}/\omega_{0,j}^2$ with \rev{$N(0)$ being the density of states at the Fermi energy}~\cite{PBAllen}. 
The asymmetric coefficient, $\omega_a$ is a phenomenological fitting parameter accounting for the fact that the Fermi level may not lie exactly halfway between the electron and hole states, as shown in Fig.~\ref{LRG-Raman}(f). The electron-phonon induced linewidth is typically negligibly small compared to the anharmonic contribution.
This is because $q=0$ phonons generally only promote vertical interband transitions to conserve both energy and momentum, and therefore have very limited phase space to excite electron-hole pairs.


We have fitted the temperature dependence of each phonon mode in \LRG\ to Eq.~\ref{ahmW} and the results are tabulated in Table~\ref{RamanTable}.
The three highest energy phonons (including $A_1^{(2)}$ and $A_1^{(3)}$ in Fig.~\ref{LRG-Raman}(c)) exhibit a conventional temperature dependence and are well captured by impurity scattering and anharmonic processes alone. However, we find that the FWHM of the five lowest energy phonons deviates from the prediction of the anharmonic decay model, and could only be well described by adding the electron-phonon interaction term.
The most extreme disagreement occurs for the $A_1^{(1)}$ phonon whose width increases markedly with decreasing temperature (Fig.~\ref{LRG-Raman}(c)).
Similar effects have been observed in graphite and graphene~\cite{Bonini2007PRL,liu2019Carbon}, and more recently in WP$_2$ and NbGe$_2$ as a signature of the phonon drag effect ~\cite{Burch2021PRX,yang2021evidence}.
\rev{Among the phonon modes of \LRG , the $A_1^{(1)}$ phonon also shows the greatest relative change in energy as a function of temperature, as can be seen in Fig.~\ref{LRG-Raman}(d).}

Unlike the line width, electron-phonon coupling has little effect on the phonon energy where the temperature dependence is contributed by two major factors: lattice thermal expansion and anharmonic phonon-phonon interaction~\cite{Menendez1984ahm}:
\begin{equation}\label{ahmE}
    \omega(T) = \omega_0 e^{-3\gamma_G \alpha_0 T} - \omega_{3p}[1+n(\omega_1,T)+n(\omega_0-\omega_1,T)].
\end{equation}
The first term accounts for the lattice thermal expansion, where $\omega_0$ is the bare phonon energy, $\gamma_G$ is the mode Gr\"{u}neisen parameter, and $\alpha_0$ is the temperature independent linear thermal expansion coefficient.
The second term arises from the same three-phonon anharmonic decay process as in Eq.~\ref{ahmW}. 
We show the change in relative energy of the $A_1$ and $E$ modes in Fig.~\ref{LRG-Raman}(d) and all of the fitting parameters are tabulated in Table \ref{RamanTable}.
All phonon energies decrease monotonically with increasing temperature, indicating they are dominated by the three-phonon anharmonic term in Eq.~\ref{ahmE}.
In some materials, it is necessary to include higher-order anharmonic terms to describe the temperature dependence of phonon self-energy~\cite{Bonini2007PRL,Mai2019}.
In \LRG, we find it only marginally improves the fit for the phonon energies and has little effect on the line width. 
This gives us confidence that the observed anomalous temperature dependence of the phonon linewidths cannot be caused by higher order anharmonic terms, and is more likely a signature of electron-phonon drag.



\section{Summary}

Our study establishes LaRhGe$_3$ as a remarkable material with many intertwined novel electronic properties. Our transport measurements  demonstrate that LaRhGe$_3$ is a compensated semimetal while our electronic structure calculations reveal Weyl nodal-line states due to the $I4mm$ space group. We report the discovery of superconductivity in \LRG\ with a \Tc\ of 0.39(1)\,K and a \mHc\ of 2.2(1)~mT. We find that \LRG\ is a type-I superconductor and well described by weak-coupling BCS theory with no evidence for TRS breaking in the zero field \muSR\ measurement suggesting a dominant singlet pairing in the measured temperature range, despite the strong antisymmetric SOC.

In the normal state, we find that the temperature dependence of resistivity in zero-field deviates from that of a typical Fermi liquid. We identify a peak in the derivative of the resistivity, the Seebeck coefficient, and in thermal conductivity at $\sim30$\,K, which is consistent with phonon drag in \LRG . We measure the temperature-dependent Raman spectra of \LRG\ and find anomalous temperature dependence of phonon line widths that can be explained by assuming electron-phonon coupling. This appears prominent in the temperature dependence of the FWHM of the lowest energy $A_1$ mode.

With the rich literature on superconductivity in materials crystallizing in the $RTX_3$ structure, we draw some comparisons with \LRG\ and find it similar to the previously reported LaRhSi$_3$.  However, our current work indicating semimetallicity and electron-phonon drag in the normal state highlights a new area of interest in the class of non-centrosymmetric materials crystallizing in the $I4mm$ space group. More work is needed to clarify the pairing state in the superconducting phase of \LRG. Furthermore, the theoretically predicted \textit{almost movable} nodal line will push further investigation of topology, both theoretically and experimentally, in the $RTX_3$ family of materials.


\section{Methods}

\subsection{Synthesis and Structural Characterization}

Single crystals of \LRG\ were grown from Rh-Ge self flux. The elements La, Rh, and Ge in the molar ratio 1:2:7 were loaded into an Al$_2$O$_3$ Canfield crucible set and sealed in a quartz tube under 0.3\,atm argon pressure. The mixture was then slowly heated to 1200\,$^\circ$C and held at this temperature for 6\,h followed by cooling to 975\,$^\circ$C over 60\,h. At 975\,$^\circ$C, the crystals were separated from the flux by centrifuging at 2000 rpm for approximately 30 s. The as-grown crystals were shiny with dimensions up to $3 \times 2 \times 1$\,mm$^3$, and are stable in air on a time scale of weeks but develop surface oxidation over more extended periods of air exposure. Phase purity and orientation of the crystals were confirmed by X-ray diffraction (XRD) using a Bruker D8 Advance with Cu K$\alpha_1$ radiation ($\lambda = 1.54056$~\AA ). The composition and homogeneity were confirmed through energy dispersive X-ray  spectroscopy (EDX) using a Philips XL30 scanning electron microscope equipped with a Bruker Quantax 200 energy-dispersion X-ray microanalysis system and an XFlash 6010 SDD detector.

\subsection{Transport and Specific Heat}

Electrical resistivity measurements were performed with conventional four- and five-probe geometries using a Quantum Design Physical Property Measurement System (PPMS). For these measurements Pt wires (25 $\mu$m) were attached to the sample with silver epoxy.
Specific heat measurements were performed using a PPMS with a $^3$He/$^4$He dilution refrigerator insert.
Thermal conductivity and Seebeck measurement were performed using a thermal transport option for the PPMS.

\subsection{\muSR\ Measurements}

Muon spin relaxation measurements were performed using a dilution refrigerator at the M15 beamline at TRIUMF. Multiple \LRG\ crystals were mounted on an Ag plate using GE varnish and placed on the sample holder. Careful canceling of the remnant field was performed using a Si semiconductor reference prior to the zero-field experiment following the method of Ref.~\cite{morris2003method}. The transverse-field experiment was performed by applying a magnetic field along the plate direction of the sample in spin-rotated mode. Muon analysis was performed in \textsc{musrfit}~\cite{suter2012musrfit}.

\subsection{Raman Measurements}


Raman scattering measurements were performed using a Horiba Jobin-Yvon LabRAM monochromator, equipped with 1800 grooves/mm grating and a Peltier cooled CCD camera.
The 632.817\,nm line of a He-Ne laser was used as the light source, and the spot diameter is about 10\,$\mu$m on the sample using a 50x microscopic objective for both focusing and collection of light. 
The spectral resolution is about 1\,cm$^{-1}$ in this study. 
A Semrock RazorEdge filter was used to block the elastically scattered laser light, where the cutoff energy is 79\,cm$^{-1}$.
For measured performed at non-ambient temperatures, we used a Cryovac Micro continuous helium flow cryostat, where the sample is thermally anchored to the copper cold-finger and sits in a vacuum environment better then $1\times 10^{-6}$\,mbar.  
Measurements were taken at temperatures ranging between 5 to 300\,K with 1\,K accuracy.

\LRG\ has five atoms in the primitive cell, leading to a total of three acoustic and 12 optical phonon bands. 
With visible light, Raman scattering only probes a small portion of the Brillouin zone near $q=0$, where the symmetries of the 12 optical phonons can be denoted by the irreducible representations of the $C_{4v}$ point group: $3A_1+ B_1+ 4E$~\cite{BilbaoRaman}.
The samples was mounted on the cryostat with the as-grown (101) surface normal to the optical axis.
At room temperature, spectra were obtained with the incident and scattered light polarized in the following combinations in order to identify the symmetries of the studied excitations:
$(E_i,E_s)=(x,x)$ couples to phonons with $A_1$, $B_1$ and $E$ symmetries, $(E_i,E_s)=(x,y)$ couples only to phonons with the $E$ symmetry.
$E_i$ and $E_s$ are the polarizations of the incident and scattered light, respectively. 
$x$ and $y$ corresponds to [10$\Bar{1}$] and [010] crystal axes, respectively.
The analyzing polarizer is removed in temperature dependent experiments to collect data from all phonons in the same cooling cycle.

\subsection{DFT Calculations}
Electronic structure calculations were performed within the framework of density functional theory (DFT) as implemented in the VASP package~\cite{vasp1, vasp2, vasp3}. 
The generalized gradient approximation with the PBE parameterization for the exchange-correlation functional~\cite{perdew1996generalized, PBE} was used. The plane-wave-basis set has been cut off at $E_{\text{cutoff}} = 287.6$ eV and the charge density was converged on a grid of 1792 irreducible $k$~points. The tetragonal crystal structure ($I4mm$, space group no. 107, \textit{a} = 4.432~\AA, and \textit{c} = 10.101~\AA) was used~\cite{kawai2008split}, and structural relaxation resulted in \textit{a} = 4.437~\AA, and \textit{c} = 10.281~\AA. The phonon modes were computed in a $2\times2\times1$ conventional supercell using the phonopy package\cite{phonopy, phonopy-phono3py-JPSJ}.

\section{Data Availability}
The data that support the findings of this study are available from the corresponding authors upon reasonable request.

\section{Author information}
\subsection{Contributions}
Single crystals were grown by S.S., M.O., and A.M.H. Heat capacity and electrical transport measurements and analysis were performed by M.O. Room temperature Raman measurements were performed by H.H.K. and M.O., and low temperature Raman measurements were performed by A.S. and B.K. Analysis of Raman data were performed by H.H.K. Thermal transport and Seebeck measurements were performed by K.P. and H.T. 
DFT calculations were performed by N.H. and A.P.S.
Muon measurements and analysis were performed by M.V.T.S., Y.C., M.O., and K.K. Manuscript was written by M.O., H.H.K., N.H., D.A.B., and A.M.H., with feedback from other co-authors.

\section{Ethics declarations}
\subsection{Competing interests}
the Authors declare no Competing Financial or Non-Financial Interests


\begin{acknowledgements}

This research was undertaken thanks in part to funding from the Max Planck-UBC-UTokyo Centre for Quantum Materials and the Canada First Research Excellence Fund, Quantum Materials and Future Technologies Program. This work was also supported by the Natural Sciences and Engineering Research Council of Canada (NSERC), the CIFAR Azrieli Global Scholars program, and the Sloan Research Fellowships program.
We thank TRIUMF staff for their technical support during muon experiment.
We thank K.~Behnia, E.~Benckiser, M.~Minola, S.~Dierker, and M.~Sigrist for valuable discussion.

\end{acknowledgements}

%


\bibliography{LaRhGe3}

\clearpage
\widetext
\begin{center}
\textbf{\large Supplemental Materials: \\Discovery of Superconductivity and Electron-Phonon Drag in \\the Non-Centrosymmetric Semimetal \LRG }
\end{center}

\setcounter{equation}{0}
\setcounter{figure}{0}
\setcounter{table}{0}
\setcounter{page}{1}
\makeatletter
\renewcommand{\theequation}{S\arabic{equation}}
\renewcommand{\thefigure}{S\arabic{figure}}
\renewcommand{\thetable}{S\arabic{table}}
\renewcommand{\bibnumfmt}[1]{[S#1]}
\renewcommand{\citenumfont}[1]{S#1}


\subsection*{Supporting details on the crystal structure of \LRG}

\begin{table}[h]
\setlength{\extrarowheight}{4pt}
\caption{\LRG\ crystals structure refinement fit results.}
\label{RefinementSummary}
\begin{tabular}{|c c|}
\hline
\textbf{Parameter}        & \textbf{Value  }    \\ \hline
Chemical Formula & LaRhGe$_3$    \\ 
Formula weight   & 459.73     \\ 
Temperature (K)      & 300       \\ 
Crystal System   & Tetragonal \\ 
Space Group      & $I4mm$        \\ 
$a$, (~\AA )           & 4.4187(2)  \\ 
$c$, (~\AA )           & 10.0494(5) \\ 
Volume (~\AA $^3$)      & 196.21(6)  \\ 
Z                &      2      \\ 
$\rho _{calc}$ (g/cm$^3$)  & 7.781      \\ 
Radiation        & Cu$_{k\alpha1}=1.5406$~\AA \\ 
$R_{\rm{Bragg}}$         & 1.544      \\ \hline
\end{tabular}
\end{table}

\begin{table}[h]
\centering
\caption{Refined atomic positions of \LRG .}
\label{RefinementAtoms}
\setlength{\extrarowheight}{8pt}
\begin{tabular}{|c c c c|}
\hline
\textbf{Atom} & \textit{x}      & \textit{y} & \textit{z}  \\
\hline
La      & 0.0 & 0.0 & 0.0\\
Rh       & 0.0 & 0.0 & 0.65762    \\
Ge1   & 0.0 & 0.0 & 0.42106  \\
Ge2    & 0.0 & 0.5 & 0.25990  \\
\hline
\end{tabular}
\end{table}

\subsection*{Supporting details on the normal state transport of \LRG}

\begin{figure}[h]
\centering
\includegraphics[width=\textwidth]{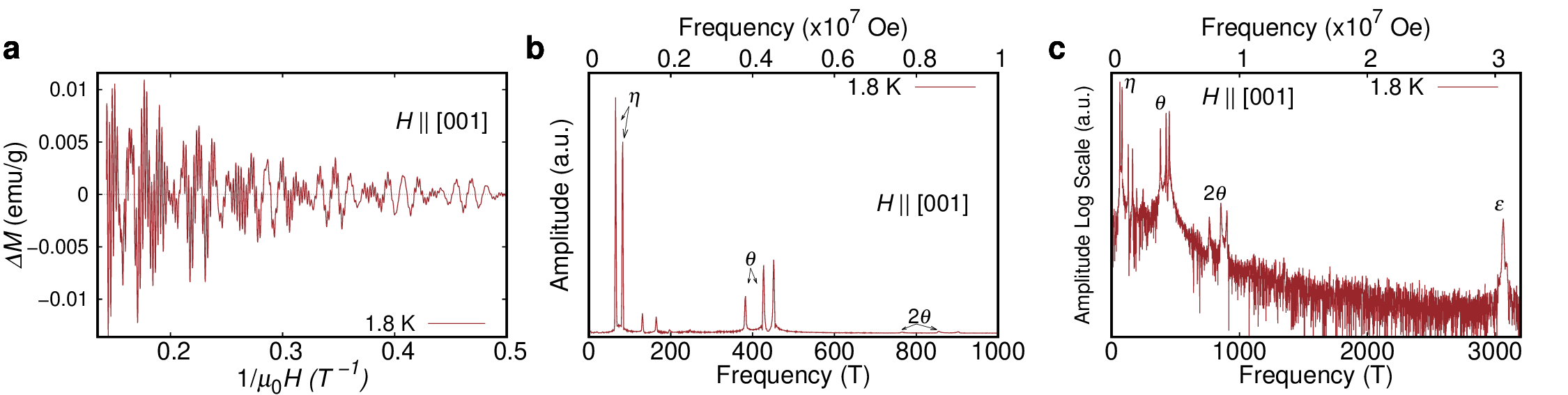}
\caption{(a) Oscillatory component $\Delta M$ corresponding to the de Haas-van Alphen (dHvA) oscillations in \LRG\ measured at 1.8\,K. (b) and (c) Fast Fourier Transform spectrum of the dHvA oscillations in (a).}
\label{LRG-dHvA}
\end{figure}

\begin{table}[h]
\centering
\caption{Comparing de Haas-van Alphen (dHvA) oscillations Frequency in our \LRG\ samples with Literature~\cite{kawai2008split}.}
\label{dHvATable}
\setlength{\extrarowheight}{8pt}
\begin{tabular}{|m{2cm} c| c|}
\hline
\textbf{Branch} & \textbf{F ($\times10^{7}$~Oe)}      & \textbf{F$_{\rm{literature}}$ ($\times10^{7}$~Oe)}~\cite{kawai2008split}  \\
\hline
$\epsilon$ (1st)        & 3.09                   & 3.06   \\
$\epsilon$ (2nd)       & 3.06                  & 3.04   \\
$\theta$ (1st)    & 0.43                  & 0.44   \\
$\theta$ (2nd)    & 0.38              & 0.38  \\
$\eta$ (1st)        & 0.082        & 0.08  \\
$\eta$ (2nd)       & 0.065        & 0.06  \\
\hline
\end{tabular}
\end{table}

\begin{figure*}[t]
\centering
\includegraphics[width=7cm]{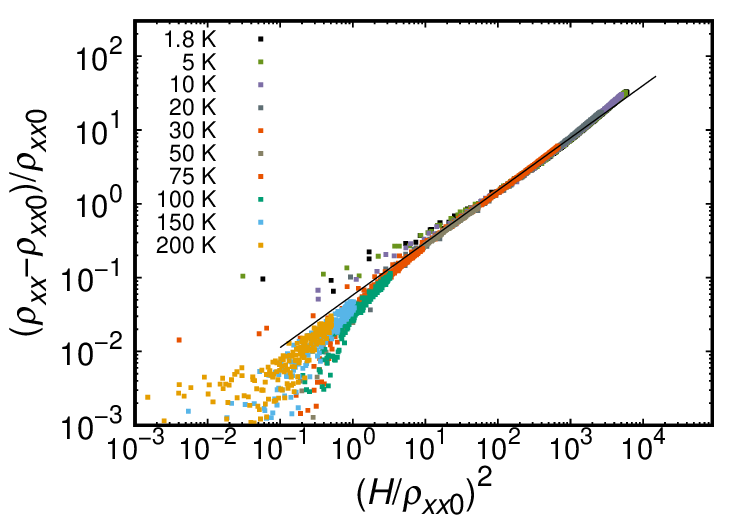}
\caption{Kohler scaling analysis of MR data shown in main text (Fig.~\ref{Fig2}). Any change of slope related to the electron phonon drag in the current data is unclear.}
\label{LRG-Kohler}
\end{figure*}

\subsection*{Supporting details on the superconductivity of \LRG}

To further characterize the superconductivity of \LRG, we next turn to modelling its specific heat. Above \Tc\ $= 0.39(1)$\,K the \Cp\ data are well described by
\begin{equation}
    C_p = \gamma _nT + \beta T^3\
    \label{CpT}
\end{equation}
\noindent
where fitting the data yields the Sommerfeld coefficient $\gamma_n = 4.92(4)$~mJ~mol$^{-1}$\,K$^{-2}$ and $\beta = 0.25(1)$~mJ~mol$^{-1}$\,K$^{-4}$. The phonon contribution is represented by the cubic term $\beta T^3$ and the Debye temperature $\Theta_D$ is then calculated according to
\begin{equation}
    \Theta_D = \left(\frac{12\pi^4}{5\beta} nR\right)^{1/3}\
    \label{ThtaD}
\end{equation}
\noindent
where $n$, the number of atoms per formula unit, is 5 and $R$ is the gas constant 8.314~J~mol$^{-1}$\,K$^{-1}$, giving $\Theta_D=341(10)\,K$. Next, taking $\Theta_D$ and \Tc, we calculate the superconducting parameter $\lambda_{ep}$ using the inverted McMillan~\cite{mcmillan1968transition} equation

\begin{equation}
    \lambda_{\text{ep}} = \frac{1.04+\mu^*ln \left( \frac{\Theta_D}{1.45T_c}\right)}{(1-0.62\mu^*) \text{ln} \left( \frac{\Theta_D}{1.45T_c}\right)-1.04}\
    \label{lamep}
\end{equation}
\noindent
where $\mu\rm{*}=0.10$ and \Tc~$= 0.39(1)$\,K. We find $\lambda_{ep}=0.34(2)$ for \LRG , suggesting it is a weak coupling superconductor. Using $\lambda_{ep}$, $\gamma$, and the Boltzmann constant $k_{\rm{B}}$, we calculate the electronic density of states (DOS) at the Fermi energy $N(E_{\rm{F}})$ with the equation

\begin{equation}
    N(E_F)=\frac{3\gamma}{\pi^2 k_B^2(1+\lambda_{\text{ep}})}\
    \label{NEf}
\end{equation}
\noindent
$N(E_{\rm{F}})$  was estimated to be 1.56 states eV$^{-1}$ per formula unit of \LRG . From our DFT calculations we find the DOS at the Fermi energy $N(E_{\rm{F}})^{calc}$ of 3.24 states eV$^{-1}$ per formula unit, and we evaluate the effective mass based on specific heat to be $m^*/m_e=N(E_{\rm{F}})/N(E_{\rm{F}})^{calc}=0.54$. This reduced effective mass compared with the free electron mass is consistent with \LRG\ being a semimetal, and the value of $m^*/m_e=0.54$ falls within the cyclotron effective masses, $m^*/m_e\sim$~0.23-1.04, reported previously by Kawai \emph{et al.}~\cite{kawai2008split}.

\begin{figure*}[t]
\centering
\includegraphics[width=18cm]{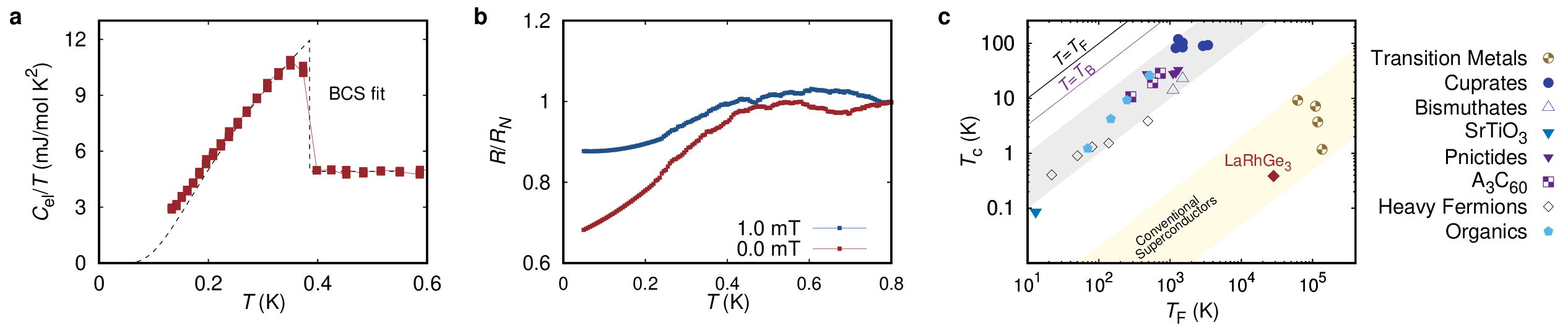}
\caption{Type-I BCS nature of the superconducting state in \LRG. (a) Superconducting transition of \LRG\ in the electronic part of the specific heat \Ce /$T$ as a function of $T$ in 0~T. The dashed black line is a fit for a \Tc\ of 0.39(1)\,K for weak-coupling BCS behavior. (b) Temperature dependence of longitudinal resistivity $\rho_\mathrm{xx}$ below 0.6~K measured in different magnetic fields, where the data are normalized by the 2.0~K data ($\rho_\mathrm{xx,2~\rm{K}}\sim1~\mu\Omega\rm{cm}$) measured in 0~mT. (c) Zero field \muSR\ spectra collected at 539~mK and 20~mK, with fit using Kubo-Toyabe function for the data. No significant change is seen between the data above and below \Tc . Inset shows the resulting relaxation rate $\sigma$. (d) Uemura plot showing the superconducting transition temperature (\Tc ) vs the Fermi temperature ($T_{\rm{F}}$) for various superconductors~\cite{uemura2004condensation,poole2007superconductivity,carbotte1990properties,peabody1972magnetic,adroja2018multigap,albedah2017absence,bhattacharyya2018brief,sonier2007hole,uemura1991muon,talantsev2015universal,uemura1989universal,lin2013fermi}, where \LRG\ falls within the region of conventional superconductors. $T=T_{\rm{B}}$ line in purple represents the Bose-Einstein condensation temperature of an ideal three-dimensional boson gas, and $T=T_{\rm{F}}$ line is shown in black. }
\label{LRG-type-I}
\end{figure*}

The magnitude of the specific heat jump $\Delta C/C_{en} \sim 1.43$, $C_{en}$ being the electronic specific heat in the normal state,  is consistent with that expected for a BCS model in the weak-coupling limit. The BCS model, shown in black in Fig.~\ref{LRG-type-I}(a), fits the data well just below \Tc , but some deviations appears at lower temperatures. With the current \Cp\ we cannot conclude whether the gap structure of \LRG\ is an isotropic $s$-wave gap or whether the gap structure has some nodes, and this will be the subject of future experiments going to lower temperatures. 

\rev{In Table~\ref{I4mm-Comparison-Table}, we compare the \Tc , $\gamma_n$, and \mHc\ values for various superconductors in the $I4mm$ space group, including \LRG\ and $\rm{La}\rm{Rh}\rm{Si}_3$. Despite the comparable $\gamma_n$ values for all of these materials, we have varying superconducting \Tc\ and critical fields. Whether this difference is due to details of electronic band structure or atomic size effects is unclear at the moment.}

\rev{We evaluate the Fermi temperature $T_{\rm{F}}$ for a 3D system via the relation $T_F=(\hbar^2/2)(3\pi^2)^{2/3}n^{2/3}/k_Bm^*$~\cite{uemura2004condensation}, where $n$ is the density of quasiparticles per unit volume.  We use the low temperature carrier density from our Hall measurements and the effective mass estimated from specific heat to obtain $T_{\rm{F}}=28340$\,K and $T_{\rm{c}}/T_{\rm{F}}=1.36\times10^{-5}$, which places \LRG\ in a region with other conventional superconductors (Fig.~\ref{LRG-type-I}(c)). This further confirms that the ternary compound \LRG\ can be described in the bulk as a BCS superconductor in the weak-coupling limit.}

\begin{table*}
\centering
\caption{Superconductivity in \LRG\ compared with other superconductors in $I4mm$ space group.}
\label{I4mm-Comparison-Table}
\setlength{\extrarowheight}{3pt}
\begin{tabular}{>{\centering}m{4cm} >{\centering}m{1.5cm}  >{\centering}m{4cm} >{\centering}m{3cm}  c}
\hline
\hline
Compound    & \Tc\ (K)  & $\gamma_n$ (mJ mol$^{-1}$ K$^{-2}$)   & \mHct\ (T) & Reference     \\ \hline
Ba(Pt,Pd)Si$_3$       & 2.3–2.8 & 4.9–5.7                             & 0.05–0.10   & \cite{bauer2009baptsi,kneidinger2014superconductivity} \\ 
La(Rh,Pt Pd,Ir)Si$_3$ & 0.7–2.7 & 4.4–6                               & Type I/0.053 & \cite{anand2011specific,kitagawa1997low,smidman2014investigations,okuda2007magnetic,anand2014physical} \\ 
Ca(Pt,Ir)Si$_3$       & 2.3–3.6 & 4.0–5.8                             & 0.15–0.27 & \cite{eguchi2011crystallographic,singh2014probing}   \\ 
Sr(Ni,Pd,Pt)Si$_3$     & 1.0–3.0 & 3.9–5.3                             & 0.039–0.174 & \cite{kneidinger2014superconductivity} \\ 
Sr(Pd,Pt)Ge$_3$        & 1.0–1.5 & 4.0–5.0                             & 0.03–0.05 & \cite{kneidinger2014superconductivity,miliyanchuk2011platinum}   \\ 
LaRhGe$_3$             & 0.39(1)   & 4.92(4)                               & Type I/0.0021(1)  & This work      \\ \hline
\hline

\end{tabular}
\end{table*}

\begin{figure*}[t]
\centering
\includegraphics[width=7cm]{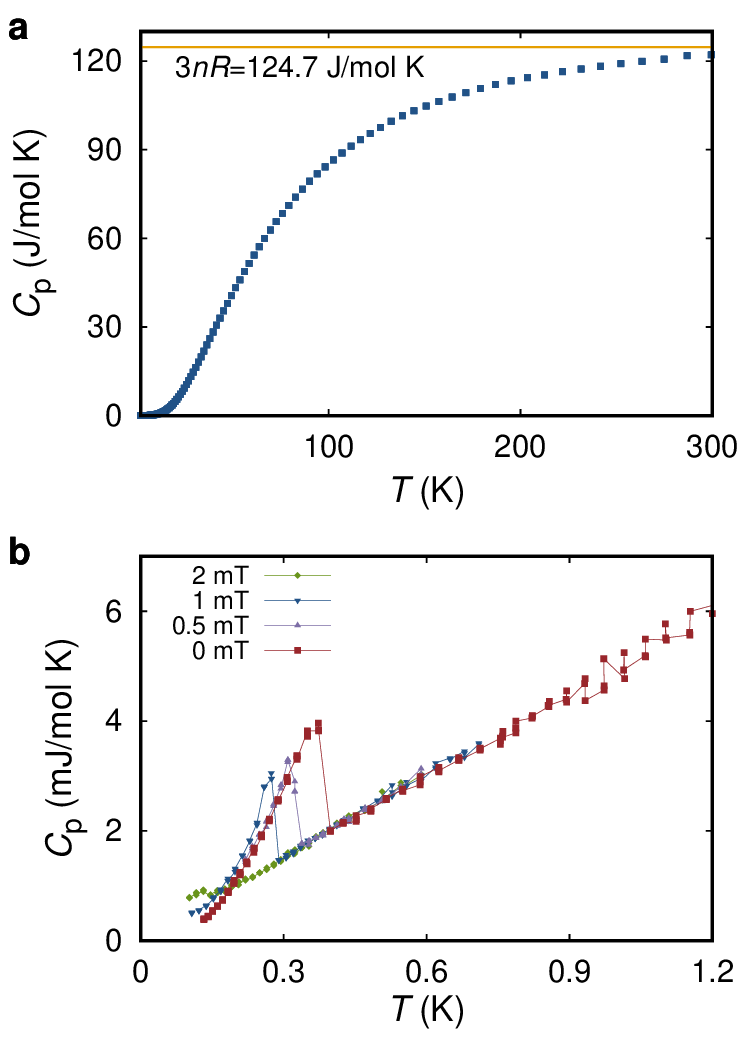}
\caption{(a) Temperature dependence of \Cp\ measured between 2-300\,K, which shows that the high temperature region approaches the Dulong-Petit limit. (b) Temperature dependence of \Cp\ measured in different applied magnetic fields, showing the superconducting transition, \Tc $0.39(1)$\,K in zero-field, that gets suppressed with applied field. Fit to the data above \Tc\ to the form $\gamma$ and $\beta$ taking into account the electronic and the phononic contributions. The phononic contribution was subtracted to obtain the \Ce\ in the main text.}
\label{LRG-Cp}
\end{figure*}

\begin{figure*}[t]
\centering
\includegraphics[width=9cm]{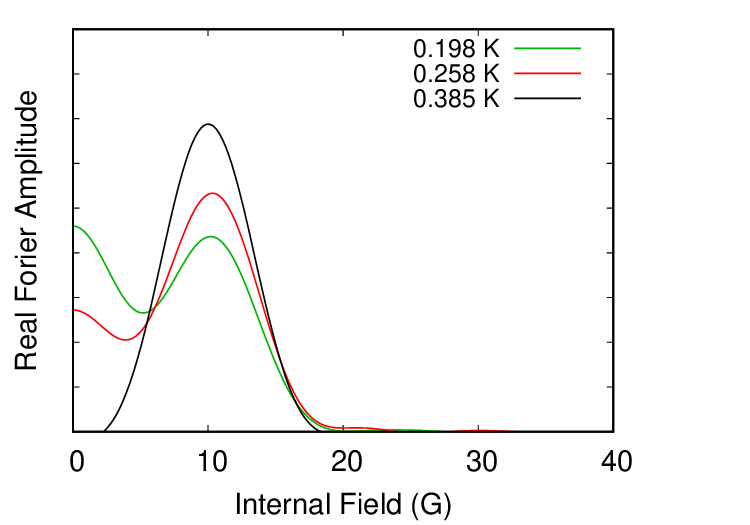}
\caption{ The Fourier transform of \muSR\ asymmetry for different temperatures in an applied field of 10~G (1~mT). A shift in the amplitude from the peak around 10~G to zero field is observed as we enter the superconducting state. No significant increase in the field of the oscillating component can be resolved from this data collected in low magnetic field.}
\label{LRG-muSR-FFT}
\end{figure*}

\begin{figure*}[t]
\centering
\includegraphics[width=12.5cm]{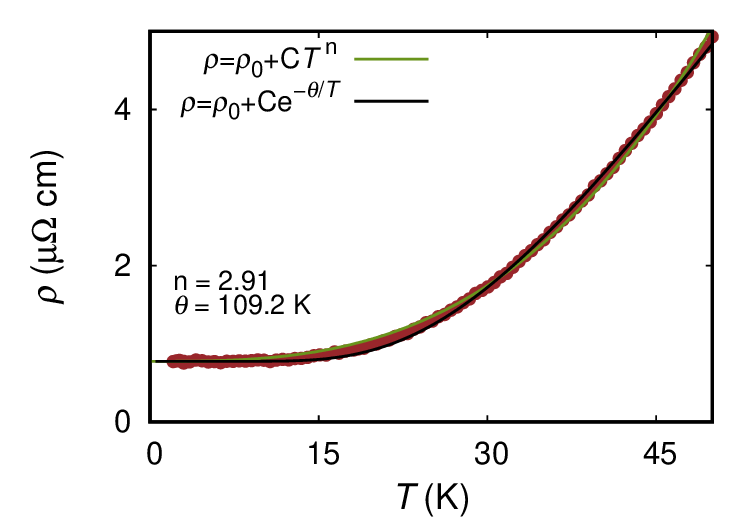}
\caption{Resistivity as a function of temperature measured along $ab$-plane of \LRG\ in zero-field, where data is fitted to exponential form of a phonon-drag model, $\rho=\rho_0+Ce^{-T_0/T}$ and power law fit $\rho=\rho_0+CT^{n}$ below 50~K.}
\label{LRG-Tn}
\end{figure*}

\clearpage

\subsection*{Supporting details on the Raman measurements of \LRG}

\begin{figure*}[h]
\centering
\includegraphics[width=7cm]{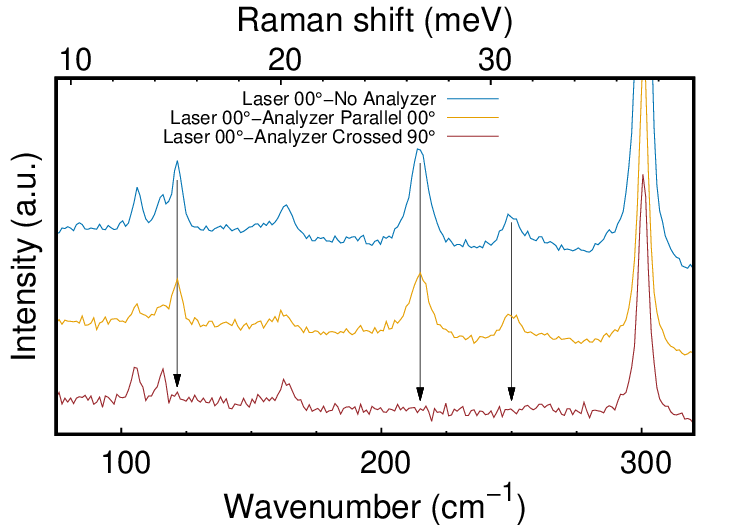}
\caption{Raman spectra collected with fixed linearly polarized laser, and polarization analyzer at different orientations. The blue curve shows data without the analyzer (mixed parallel and perpendicular); orange curve is the analyzer polarization parallel to the laser (xx geometry); red curve is the analyzer 90$^{\circ}$ rotated with respect to laser polarization (xy geometry). Three phonon modes that are strongly suppressed were highlighted with arrows, demonstrating that these are $A_1$ modes of \LRG.}
\label{LRG-RamanPolarization}
\end{figure*}

\begin{figure}[t]
\centering
\includegraphics[width=7cm]{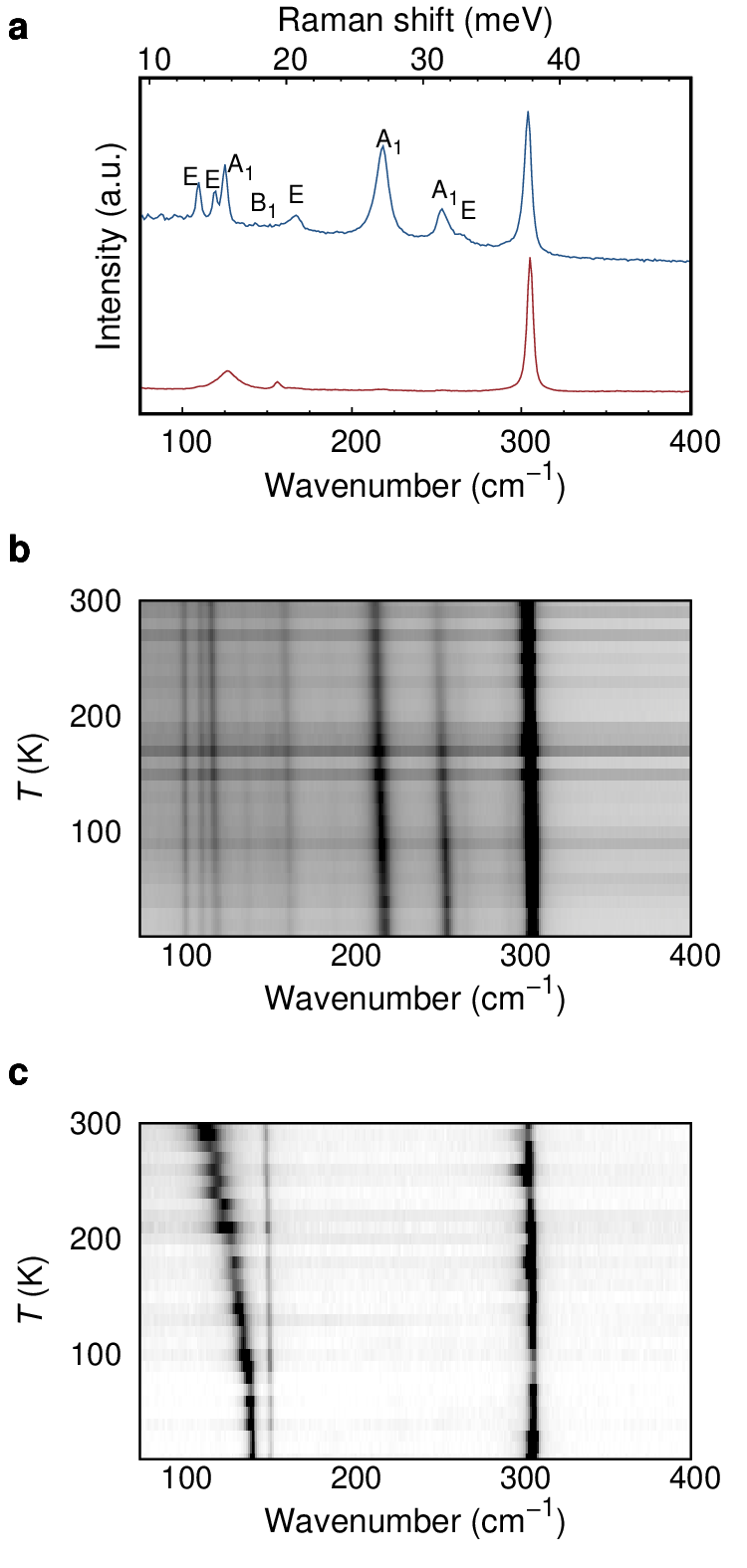}
\caption{(a) Raman spectra of \LRG\ measured along different directions of the crystal at room temperature. (b) Black and white map of Raman scattering intensity plotted as a function of temperature and Raman shift measured from the (101) surface of \LRG\ crystal. Unless otherwise specified, all of the Raman data in this work were taken from this sample.
(c) Black and white map of Raman scattering intensity measured from the same sample as panel (b), but from a different spot. \rev{Since these data were not reproduced by other crystals that we measured}, this spot is likely a crystallite with different orientation and possibly even different stoichiometry as the host crystal. However, we highlight the mode at 144~cm$^{-1}$ at 10\,K where the energy softens drastically with increasing temperature, to 120~cm$^{-1}$ at 300\,K. Such large temperature dependence is highly unusual and is difficult to reconcile with the moderate anharmonicity found in \LRG. The energy of this mode is in the vicinity of the the $A_1^{(1)}$ mode that shows strong electron-phonon coupling, raising the possibility of electron-phonon induced self-energy effect. However, more study is required on this crystallite before any conclusion can be drawn. In this paper, our discussion is focused on data in panel (b) which is reproduced across samples.}
\label{LRG-Raman-Sup}
\end{figure}

\begin{figure}[t]
\centering
\includegraphics[width=7cm]{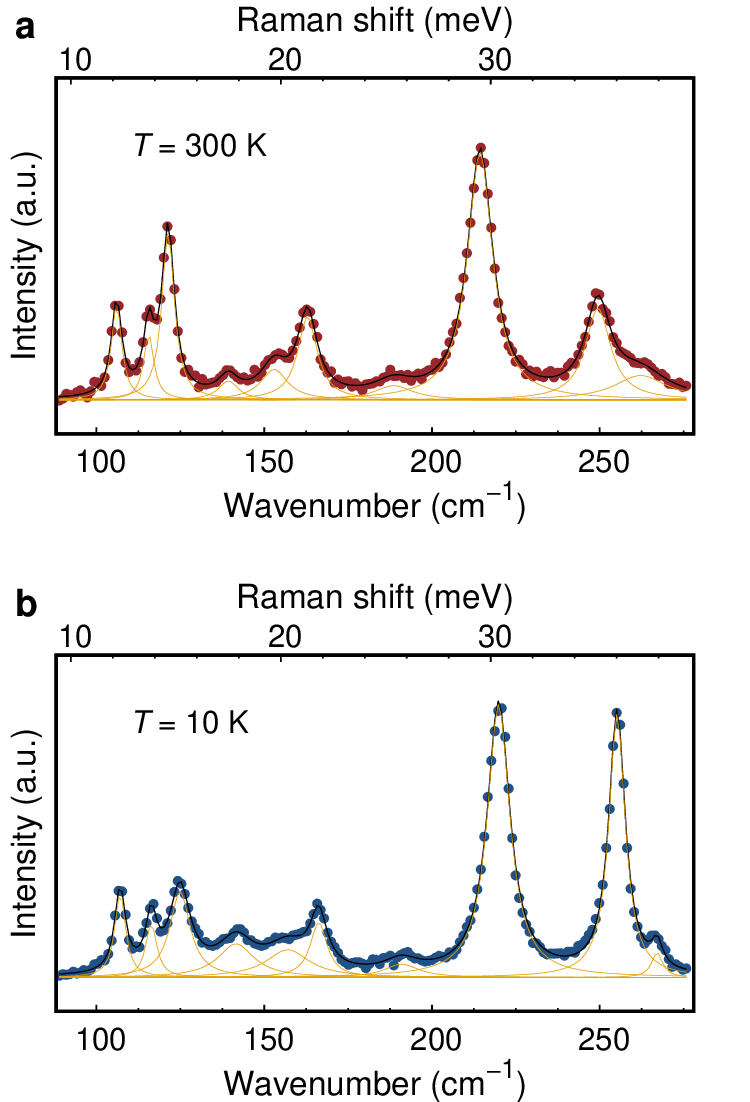}
\caption{Raman spectra of \LRG\ measured at 300 K (a) and 10 K (b), sliced from Fig.~\ref{LRG-Raman-Sup}(b) to showcase the extraction of phonon energy and width through fitting. Measured data shown as points and fitted curves for each phonon mode shown individually in yellow and total fit in black. }
\label{LRG-Raman-Fit-Sup}
\end{figure}

\begin{table}[h]
\centering
\caption{Calculated energies and symmetries of the phonon modes for \LRG\ and comparison with the fitting parameters of Eq.~\ref{ahmE} based on our Raman spectra in Fig.~\ref{LRG-Raman-Sup}. All units are in cm$^{-1}$. The errors represent one standard deviation of the least-squared fit.}
\label{RamanTable}
\setlength{\extrarowheight}{8pt}
\begin{tabular}{|m{2cm}m{2cm}lllllll|}
\hline
\textbf{DFT} & \textbf{$\omega$(10\,K)} & \textbf{$\omega_0$} & \textbf{$\omega_1$} & \textbf{$\omega_{3p}$} & \textbf{$\gamma_0$} & \textbf{$\gamma_{3p}$} & \textbf{$\gamma_{ep}$} & \textbf{$\omega_a$}\\ \hline
$E^{(1)}$ 98.1 & 107.0(1) & 107.3(1) & $\omega_0/2$ & 0.18(8) & 1.8(4) & 0.27(6) & 2.3(7) & 47(4)     \\
$E^{(2)}$ 108.6 & 116.3(1) & 116.6(1) & $\omega_0/2$ & 0.14(19) & 1.6(5) & 0.36(8) & 2.0(9) & 39(19)     \\
$A_1^{(1)}$ 112.8 & 124.9(1) & 125.2(1) & $\omega_0/2$ & 0.60(2) & 0.9(8) & 0.59(14) & 9.9(15) & 61(1)     \\
$B_1^{(1)}$ 159.1 & 141.6(6) & 142.6(4) & $\omega_0/2$ & 0.50(60) & 0(17) & 1.0(30) & 24(29) & 71(5)     \\
$E^{(3)}$ 162.0 & 166.2(2) & 167.2(1) & $\omega_0/2$ & 0.86(2) & 3(10) & 0.7(18) & 2.1(77) & -14(645)     \\
$A_1^{(2)}$ 207.9 & 219.9(0) & 221.1(1) & $\omega_0/2$ & 1.85(5) & 8.2(1) & 0.22(3) & -- & --     \\
$A_1^{(3)}$ 240.8 & 255.2(0) & 256.6(4) & 52(6)        & 1.4(4) & 5.1(1) & 0.84(4) & -- & --     \\
$E^{(4)}$ 252.7 & 267.1(3) & 268.8(15) & 61(22)      & 1.6(15) & 1.4(9) & 2.4(3) & -- & --     \\
\hline
\end{tabular}
\end{table}





\end{document}